\documentclass{article}
\usepackage{color,epic,eepicemu,epsfig}

\definecolor{gray}{gray}{0.7}

\newcommand{\pmt}{%
\makebox(0,0){\color{gray}\scalebox{0.7}{%
\path(-1.2,+0.0)(-1.2,+0.5)%
\qbezier(-1.2,+0.5)(0.1,+0.5)(0.3,+1.0)%
\qbezier( 0.3,+1.0)(0.7,+2.0)(1.1,+1.0)%
\qbezier( 1.1,+1.0)(1.5, 0.0)(1.1,-1.0)%
\qbezier( 0.3,-1.0)(0.7,-2.0)(1.1,-1.0)%
\qbezier(-1.2,-0.5)(0.1,-0.5)(0.3,-1.0)%
\path(-1.2,-0.0)(-1.2,-0.5)%
\color{black}}}%
}

\begin{document}
\raggedright
\sffamily

\title{The probability density function\\ of\\ the arrival time of \v{C}erenkov light}
\author{M.\ de Jong\\
  {\small NWO-I, Nikhef, PO Box 41882, Amsterdam, 1098 DB Netherlands} \\
  {\small Leiden University, Leiden Institute of Physics, PO Box 9504, Leiden, 2300 RA Netherlands}\\
  E. van Campenhout\\
  {\small Leiden University, Leiden Institute of Physics, PO Box 9504, Leiden, 2300 RA Netherlands}\\
}
\maketitle

\begin{abstract}
The probability density function 
of the arrival time of \v{C}erenkov light on a photo-multiplier tube 
has been studied.
This study covers light production, transmission and detection.
The light production includes 
the light from a muon,
the light from a shower and
the light due to the energy loss of a muon.
For the transmission of light, the effects of 
dispersion, 
absorption and 
scattering in the medium are considered.
For the detection of light, 
the angular acceptance and
the quantum efficiency of the photo-multiplier tube are taken into account.
\end{abstract}

\section{Introduction}
\label{Introduction}

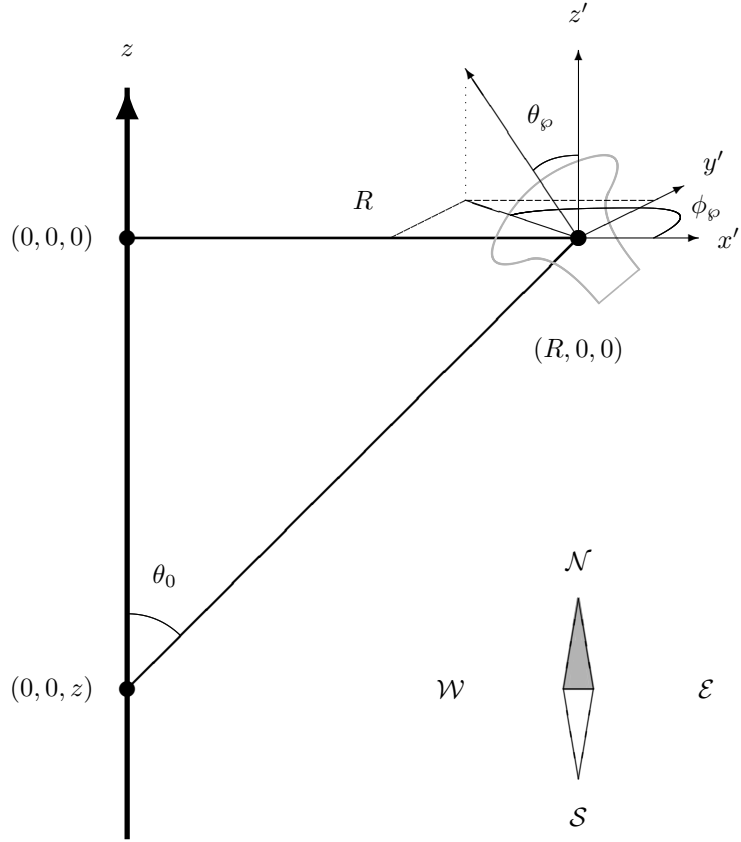
\begin{figure}[ht!]
\setlength{\unitlength}{1cm}
\begin{center}
\begin{picture}(10,12)
\allinethickness{0.35mm}%
\put(1,0){\scalebox{1}{%
\put( 1.0, 0.0){\scalebox{2}{\vector(0,1){5}}}%
\put( 1.0,10.5){\makebox(0,0){$z$}}%
\put( 1.0, 8.0){\line(1,0){6}}%
\put( 4.0, 8.5){\makebox(0,0)[l]{$R$}}%
\put( 1.0, 2.0){\line(1,1){6}}%
\put( 7.0, 8.0){\rotatebox{+130}{\scalebox{1}{\pmt}}}%
\put( 1.0, 8.0){\circle*{0.2}}%
\put( 0.0, 8.0){\makebox(0,0){$(0,0,0)$}}%
\put( 7.0, 8.0){\circle*{0.2}}%
\put( 7.0, 6.5){\makebox(0,0){$(R,0,0)$}}%
\put( 1.0, 2.0){\circle*{0.2}}%
\put( 0.0, 2.0){\makebox(0,0){$(0,0,z)$}}%
\thinlines%
\put( 1.0, 2.0){\qbezier(0,1)(0.41,1)(0.71,0.71)}%
\put( 1.5, 3.5){\makebox(0,0){$\theta_0$}}%
\put( 7.0, 8.0){%
\put(0,0){\vector(0,1){2.5}}%
\put(0,0){\vector(1,0){1.6}}%
\put(0,0){\vector(2,1){1.4}}%
\put(0.0,3.0){\makebox(0,0){$z'$}}%
\put(2.0,0.0){\makebox(0,0){$x'$}}%
\put(1.8,1.0){\makebox(0,0){$y'$}}%
\put(0,0){\vector(-2,3){1.5}}%
\dashline[100]{0.1}(-2.5,0.0)(-1.5,0.5)(1.0,0.5)%
\dottedline{0.1}(-1.5,0.5)(-1.5,2.0)%
\put(0,0){\line(-3,1){1.5}}%
\put( 0.0,0.0){\circle*{0.2}}%
\put( 0.0,0.0){\qbezier(1,0)(1.8,0.4)(0.8,0.4)}%
\put( 0.0,0.0){\qbezier(0.8,0.4)(-0.6,0.4)(-0.9,0.3)}%
\put( 1.7,0.4){\makebox(0,0){$\phi_\wp$}}%
\put( 0.0,0.0){\qbezier(0,1.1)(-0.4,1.1)(-0.6,0.9)}%
\put(-0.5,1.6){\makebox(0,0){$\theta_\wp$}}%
}%
\put(7,2){%
\put( 0.0,+1.7){\makebox(0,0){$\mathcal{N}$}}%
\put(+1.7, 0.0){\makebox(0,0){$\mathcal{E}$}}%
\put( 0.0,-1.7){\makebox(0,0){$\mathcal{S}$}}%
\put(-1.7, 0.0){\makebox(0,0){$\mathcal{W}$}}%
\put( 0.0, 0.0){\color{gray}\allinethickness{0.5mm}%
\path(-0.20,0.0)(+0.20,0.0)%
\path(-0.20,0.0)(0.0,+1.2)(+0.20,0.0)%
\path(-0.18,0.0)(0.0,+1.2)(+0.18,0.0)%
\path(-0.16,0.0)(0.0,+1.2)(+0.16,0.0)%
\path(-0.14,0.0)(0.0,+1.2)(+0.14,0.0)%
\path(-0.12,0.0)(0.0,+1.2)(+0.12,0.0)%
\path(-0.10,0.0)(0.0,+1.2)(+0.10,0.0)%
\path(-0.08,0.0)(0.0,+1.2)(+0.08,0.0)%
\path(-0.06,0.0)(0.0,+1.2)(+0.06,0.0)%
\path(-0.04,0.0)(0.0,+1.2)(+0.04,0.0)%
\path(-0.02,0.0)(0.0,+1.2)(+0.02,0.0)%
\path(-0.00,0.0)(0.0,+1.2)(+0.00,0.0)%
\thinlines\color{black}}%
\path(-0.2,0.0)(0.0,-1.2)(+0.2,0.0)%
\path(-0.20,0.0)(+0.20,0.0)%
\path(-0.2,0.0)(0.0,+1.2)(+0.2,0.0)%
}%
}}%
\end{picture}
\end{center}
\caption{
Topology of a muon or shower producing light that is detected on a PMT.
The muon or shower direction is pointed along the $z-$axis and the PMT is located at position $(R,0,0)$.
The zenith and azimuth angle of the orientation of the PMT are denoted by 
$\theta_{\wp}$ and $\phi_{\wp}$, respectively.
The compass refers to the orientation of the PMT when 
its axis lies within the $x-z$ plane (i.e. $\sin\phi_\wp = 0$).
\label{f:event}}%
\end{figure}

The generic topology of a muon or a shower producing light that is detected on a photo-multiplier tube (PMT) 
is shown in figure \ref{f:event}.
The coordinate system is defined such that 
the muon or shower direction is pointed along the $z-$axis and
the position of the PMT is located in the $x-z$ plane.
The orientation of the PMT is defined by 
the zenith angle, $\theta_{\wp}$, and 
the azimuth angle, $\phi_{\wp}$.
The distance of closest approach of the muon or shower to the PMT is denoted by $R$.
This topology is obtained after the following operations:

\begin{enumerate}
\item Rotation \\
  -- Muon or shower direction is pointed along the $z-$axis.
\item Translation \\
  -- Extrapolation of muon or shower trajectory passes through the coordinate origin.
\item Rotation \\
  -- PMT is located in the $x-z$ plane.
\end{enumerate}

The rotation of the coordinate system, $\mathcal{R}$, can be expressed  
as a $3\times3$ matrix:

\begin{eqnarray}
  \mathcal{R} = 
  \left( 
  \begin{array}{ccc}
      \cos\theta \, \cos\phi &  \cos\theta \, \sin\phi &  -\sin\theta \\
              -\sin\phi      &          \cos\phi       &       0      \\
      \sin\theta \, \cos\phi &  \sin\theta \, \sin\phi &  +\cos\theta \\
    
  \end{array}
  \right)       \label{eq:rotation}
\end{eqnarray}

For the first rotation, 
$\theta$ and $\phi$ correspond to 
the zenith and
the azimuth angle
of the direction of the muon or shower in the original frame, respectively.
The $x$ and $y$ values of the position of the muon or shower in the rotated system (i.e.\ after step 1) 
are then used to translate the coordinate system such that 
the extrapolation of the muon or shower trajectory passes through the origin.
A second rotation is applied to the coordinate system 
such that the position of the PMT is located in the $x-z$ plane.
For this rotation, $\theta = 0$ and $\phi = \textrm{atan2}(y,x)$,
where $x$ and $y$ refer to the position of the PMT after step 2.
\newline

\begin{figure}[ht!]
\setlength{\unitlength}{1cm}
\begin{center}
\begin{picture}(14,6)
\allinethickness{0.35mm}%
\put(0.0,0.0){\scalebox{1}{%
\put( 1.0, 0.0){\vector(0,1){5}}%
\put( 1.0, 5.5){\makebox(0,0){$z$}}%
\put( 1.0, 4.0){\vector(1,0){4}}%
\put( 5.5, 4.0){\makebox(0,0){$x$}}%
\put( 1.0, 1.0){\path(0,0)(2.0,0.5)(3,3)}%
\put( 2.0, 0.7){\makebox(0,0){$\bar{u}$}}%
\put( 4.0, 2.7){\makebox(0,0){$\bar{v}$}}%
\put( 1.0, 4.0){\circle{0.2}}%
\put( 1.0, 4.0){\makebox(0,0){$\otimes$}}%
\put( 0.3, 4.0){\makebox(0,0){$y$}}%
\put( 4.0, 4.0){\circle*{0.2}}%
\put( 4.0, 4.5){\makebox(0,0){$(R,0)$}}%
\put( 1.0, 1.0){\circle*{0.2}}%
\put( 0.3, 1.0){\makebox(0,0){$(0,z)$}}%
\thinlines%
\put( 1.0, 1.0){\qbezier(-0.00,+1.00)(+0.75,+1.00)(+0.96,+0.28)}%
\put( 1.7, 2.3){\makebox(0,0){$\theta_0'$}}%
}}%
\put(7.0,1.0){\scalebox{1}{%
\put( 1.0, 0.0){\vector(1,0){5}}%
\put( 6.5, 0.0){\makebox(0,0){$x$}}%
\put( 1.0, 0.0){\vector(0,1){2.5}}%
\put( 1.0, 3.0){\makebox(0,0){$y$}}%
\put( 1.0, 0.0){\path(0,0)(3,+2)(4,0)}%
\put( 2.2, 1.5){\makebox(0,0){$\bar{u}$}}%
\put( 5.0, 1.3){\makebox(0,0){$\bar{v}$}}%
\put( 1.0, 0.0){\circle{0.2}}%
\put( 1.0, 0.0){\makebox(0,0){$\odot$}}%
\put( 0.3, 0.0){\makebox(0,0){$z$}}%
\put( 5.0, 0.0){\circle*{0.2}}%
\put( 5.0,-0.5){\makebox(0,0){$(R,0)$}}%
\thinlines%
\put( 1.0, 0.0){\qbezier(+0.86,+0.51)(+1.00,+0.27)(+1.00,+0.00)}%
\put( 2.5,+0.4){\makebox(0,0){$\phi_0$}}%
}}%
\end{picture}
\end{center}
\caption{
Topology of a muon or a shower producing light that is detected on a PMT after a single scattering of the photons 
(left $x-z$ view and right $x-y$ view).
The muon or shower direction is pointed along the $z-$axis and the PMT is located at position $(R,0,0)$.
The angle $\theta_0'$ is defined by $\tan\theta_0' \,=\, \frac{\sin\theta_0\cos\phi_0}{\cos\theta_0}$. 
\label{f:scattering}}%
\end{figure}
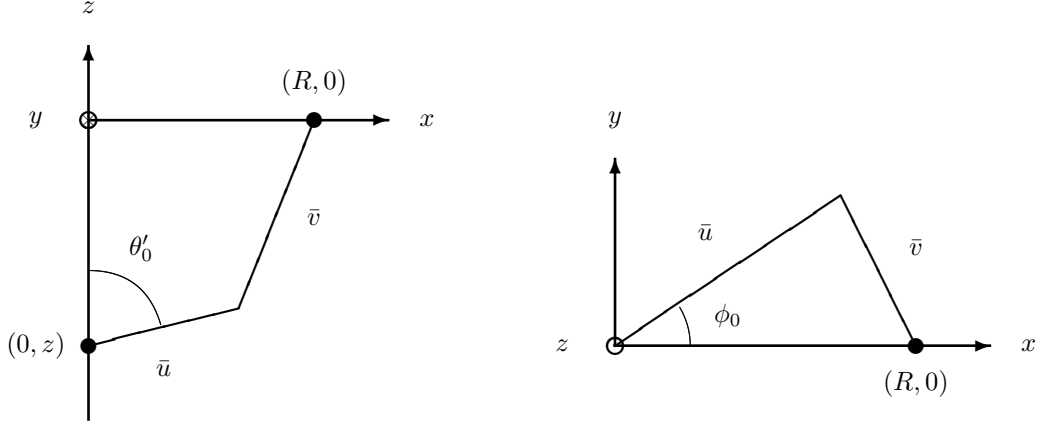

The topology of a muon producing light that is detected on a PMT after a single scattering of the light 
is shown in figure \ref{f:scattering}.
The complete path of the photon from 
a   position along the muon trajectory to 
the position of the PMT 
can be expressed as the sum of the two vectors $\bar{u}$ and $\bar{v}$.
The scattering angle, $\theta_s$, is then defined as:

\begin{eqnarray}
  \cos\theta_s & \equiv & \hat{u} \cdot \hat{v} \label{eq:scattering-angle}
\end{eqnarray}

where
$\hat{u}$ ($\hat{v}$) corresponds to the unit direction vector of the photon before (after) the scattering. 
In addition to the zenith angle, $\theta_0$, 
the azimuth angle, $\phi_0$, is required to describe the direction of the emitted photon 
(i.e.\ $\hat{u}$).
This angle is defined as the angle between 
the $x-z$       plane and
the $\hat{u}-z$ plane (see figure \ref{f:scattering}).
\newline

As a result of the coordinate transformations, 
the time response of the PMT to the various sources of light is completely determined by 
the distance between the PMT and the muon or shower and the orientation of the PMT.
In the following sections, 
the calculation of the probability density function of the arrival time of \v{C}erenkov light is explained and 
in section \ref{Example plots}, results are presented for the deep-sea water and PMT used in KM3NeT (see \verb'www.km3net.org').

\section{Light production, transmission and detection}
\label{Light production, transmission and detection}

In the following, the production, transmission and detection of light is presented.
The light production includes 
the \v{C}erenkov light from a muon,
the light from showers and
the light due to the energy loss of a muon.
For the transmission of light, the effects of 
dispersion,
absorption and 
scattering in the medium are considered.
For the detection of light, 
the angular acceptance and
the quantum efficiency of the photo-multiplier tube are taken into account.

\subsection{\v{C}erenkov light}
\label{Cherenkov light}

The number of \v{C}erenkov photons produced per unit path length of a particle 
with charge, $ze$, moving with speed, $\beta$, through a medium
can be expressed as \cite{ref:pdg}:

\begin{eqnarray}
  \frac{d^2N}{dx d\lambda} & = & \frac{2\pi\alpha z^2}{\lambda^2}\left( 1 - \frac{1}{\beta^2 n^2} \right) \label{eq:cherenkov}
\end{eqnarray}

where
$\lambda$ is the wavelength of the light,
$\alpha$ the Electro-Magnetic coupling constant, and
$n$ the index of refraction of the medium.
The index of refraction of the medium and 
the characteristic angle $\theta_0 = \theta_C$ of the \v{C}erenkov cone 
are related in the following way:

\begin{eqnarray}
  \cos\theta_C & = & 1/n    \label{eq:cherenkov-angle}
\end{eqnarray}
where we have assumed a highly relativistic particle, such that $\beta\simeq1$.

\begin{figure}[ht!]
\setlength{\unitlength}{1cm}
\begin{center}
\begin{picture}(7,6)
\allinethickness{0.35mm}%
\put(1,0){\scalebox{1}{%
\put( 1.0, 0.00){\vector(0,1){6}}%
\put( 1.0, 5.00){\line(1,0){3.0}}%
\put( 1.0, 4.50){\line(1,0){3.5}}%
\put( 2.5, 5.50){\makebox(0,0){$R$}}%
\put( 1.0, 5.00){\line(-1,0){0.1}}%
\put( 1.0, 4.50){\line(-1,0){0.1}}%
\put( 0.5, 4.75){\makebox(0,0)[r]{$dz' = dz\,\sin^2\theta_0$}}%
\put( 1.0, 2.00){\path(0,0)(3.0,3.0)}%
\put( 1.0, 1.00){\path(0,0)(3.5,3.5)}%
\put( 1.0, 2.00){\line(-1,0){0.1}}%
\put( 1.0, 1.00){\line(-1,0){0.1}}%
\put( 0.5, 1.50){\makebox(0,0){$dz$}}%
\put( 1.5, 1.50){\path(0,0)(-0.5,+0.5)}%
\put( 4.5, 4.50){\path(0,0)(-0.5,+0.5)}%
\put( 4.0, 5.00){\path(0,0)(0.07,0.07)}%
\put( 4.5, 4.50){\path(0,0)(0.07,0.07)}%
\put( 4.7, 4.75){\makebox(0,0)[lb]{$dz\,\sin\theta_0$}}%
\put( 4.0, 4.50){\path(0,0)( 0.0,+0.5)}%
%
\put( 1.5, 3.20){\makebox(0,0){$\theta_0$}}%
}}%
\end{picture}
\end{center}
\caption{
Relation between the length of a track segment ($dz$) and the height of the light cone ($dz'$).
The area of the light cone is $A = 2 \pi R dz\,\sin\theta_0$.
\label{f:light-cone}}%
\end{figure}
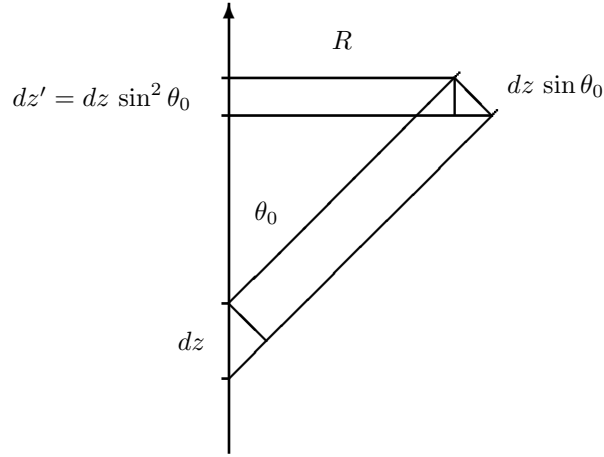

In the hypothesis of a light cone, 
the length of a track segment and the height of the light cone 
are related (see figure \ref{f:light-cone}).
The number of detectable photons per unit wavelength and per unit area at a distance, $R$, can thus be formulated as:

\begin{eqnarray}
  \Phi_0(R,\lambda) & = & \frac{d^2N}{dx d\lambda} \: \frac{1}{2 \pi R \: \sin\theta_C}  \label{eq:I0}
\end{eqnarray}

\subsection{Light from showers}
\label{Light from showers}

The detectable signal from showers is primarily due to the \v{C}erenkov light produced by charged particles in the shower.
As the hadronic interaction length and the radiation length of water (and ice) are very similar,
the light from both hadronic and Electro-Magnetic showers are treated in the same way.
It is convenient to express the number of detectable photons per unit wavelength and per unit shower energy 
in terms of 
the equivalent track length per unit shower energy and 
the number of detectable photons per unit wavelength and per unit track length (equation \ref{eq:cherenkov}), i.e:

\begin{eqnarray}
  \frac{d^2N}{dE d\lambda} & = & \frac{dx}{dE} \frac{d^2N}{dx d\lambda}                  \label{eq:I1.1}
\end{eqnarray}

The equivalent track length depends only on the medium and not on the PMT.
For water (and ice) it typically amounts to about $4.7 ~\mathrm{m/GeV}$.
\newline

The angular distribution of light emission from an Electro-Magnetic shower has been studied 
extensively and is presented in references \cite{ref:mirani, ref:copper}.
For energies in excess of $1\,\mathrm{GeV}$, 
it has been found that the angular distribution is rather 
independent of the energy of the Electro-Magnetic shower.
The angular distribution can be parametrised reasonably well as \cite{ref:kooijman}:

\begin{eqnarray}
  \frac{d^2P_\star}{d\cos\theta_0\:d\phi_0} & = & c \: e^{\textstyle b \left| \cos\theta_0 -  \cos\theta_C \right| ^a}
\end{eqnarray}

where

\begin{eqnarray*}
  a & = & +0.35                                  \\*[1mm]
  b & = & -5.40                                  \\
  c & = & \frac{1}{2\pi} \frac{1}{0.06667}       \\
\end{eqnarray*}

\begin{figure}[ht!]
\begin{center}
\begin{picture}(8,7)
\allinethickness{0.35mm}%
\put(4.2,-0.1){\makebox(0,0)[b]{\scalebox{0.4}{\includegraphics{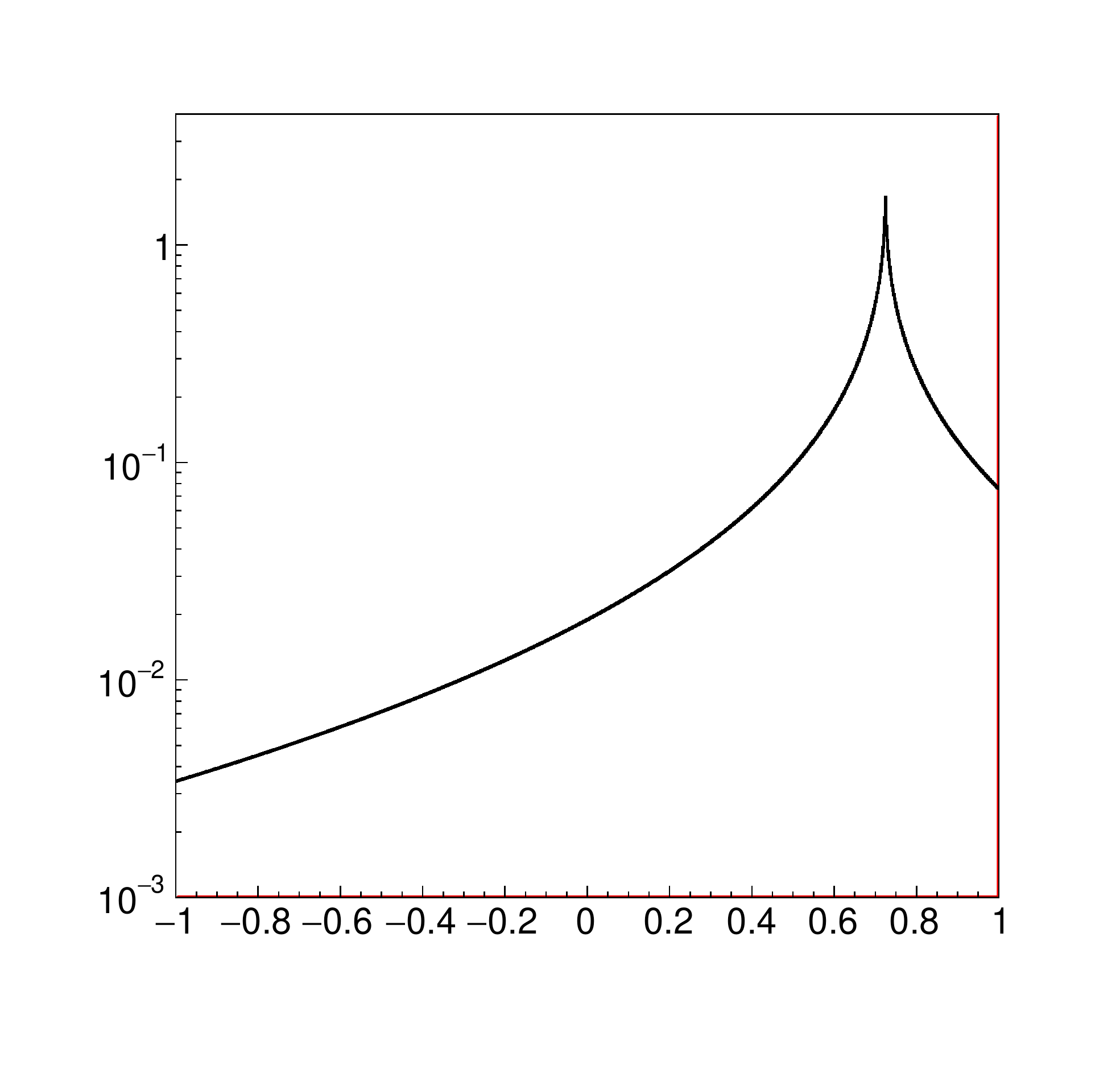}}}}%
\put(4.2,-0.2){\makebox(0,0){$\cos\theta$}}%
\put(0.0, 3.7){\makebox(0,0){\rotatebox{90}{$\frac{\textstyle d^2P_\star}{\textstyle d\cos\theta\,d\phi}$}}}%
\end{picture}
\end{center}
\caption{\label{f:geant} Parametrisation of the angular distribution of light emission from an EM-shower.}
\end{figure}

The constant, $c$, is defined such that $P_\star$ is normalised to unity 
for the full solid angle.
The result of the parametrisation is shown in figure \ref{f:geant}.

The number of detectable photons 
per unit wavelength, per unit energy and per unit solid angle
as a function of 
the angle of emission 
can then be formulated as:

\begin{eqnarray}
  \Phi_1(\cos\theta_0, \lambda) & = & 
  \frac{d^2N}{dEd\lambda} \:
  \frac{d^2P_\star}{d\cos\theta_0\:d\phi_0}                                              \label{eq:I1}
\end{eqnarray}

The longitudinal profile of a shower has been presented in reference \cite{ref:copper}.
It can be parametrised reasonably well as:

\begin{eqnarray}
  \frac{dP}{dz}  & = & z^{a-1} \frac{e^{-z/b}}{b^a \, \Gamma(a)}                         \label{eq:dP/dz}
\end{eqnarray}

where

\begin{eqnarray*}
  a & = & 1.85 \, + \, 0.62 \times \log\frac{E}{\mathrm{GeV}}   \\*[1mm]
  b & = & 0.54                                                  \\
\end{eqnarray*}

where $E$ is the energy of the shower.
The normalisation is defined such that the integral from $0$ to $\infty$ is normalised to unity.

\subsection{Light due to energy loss of a muon}
\label{Light due to energy loss of a muon}

The energy loss of the muon per unit track length can be expressed as \cite{ref:pdg}:

\begin{eqnarray}
  -\frac{dE}{dx} & = & a(E) + b(E) \: E \label{eq:energy-loss}
\end{eqnarray}

where 
$a$ refers to the ionisation energy loss and
$b$ to the sum of $e^+e^-$ pair production and bremsstrahlung.
The $e^+e^-$ pair production and bremsstrahlung both contribute to the detectable signal.
The number of detectable photons per unit wavelength and per unit track length 
due to the energy loss of a muon can then be formulated as:

\begin{eqnarray}
  \frac{d^2N(E)}{dx d\lambda} & = & b(E) \, E \, \frac{d^2N}{dE d\lambda}                \label{eq:I2.1}
\end{eqnarray}

The number of detectable photons 
per unit wavelength, per unit track length and per unit solid angle
as a function of 
the energy of the muon and
the angle of emission 
can then be formulated as:

\begin{eqnarray}
  \Phi_2(\cos\theta_0, E, \lambda) & = &
  b(E) \, E \,
  \Phi_1(\cos\theta_0, \lambda)                                                           \label{eq:I2}
\end{eqnarray}

There is also a contribution of energetic knock-on electrons ($\delta$ rays).
The energy loss due to $\delta$ rays can be expressed as \cite{ref:pdg}:

\begin{eqnarray}
  T \frac{d^2N}{dT dx} & = & \frac{1}{2} K z^2 \frac{Z}{A} \frac{1}{\beta^2} \frac{F(T)}{T}    \label{eq:I3.1}
\end{eqnarray}

where $T$ is the kinetic energy of the knocked-on electron.
The minimal and maximal kinetic energy can be expressed as:

\begin{eqnarray*}
  T_{min} & = & m_ec^2 \frac{1}{\sqrt{n^2-1}}                                                  \\
  T_{max} & = & \frac{2 m_ec^2 \beta^2 \gamma^2}{1 + 2\gamma m_e/M_{\mu} + (m_e/M_{\mu})^2}       
\end{eqnarray*}

where $\beta$ and $\gamma$ refer to the speed and the Lorentz factor of the muon, respectively.\\*[\baselineskip]

The number of detectable photons 
per unit wavelength, per unit track length and per unit solid angle
as a function of 
the energy of the muon and
the angle of emission 
can then be formulated as:

\begin{eqnarray}
  \Phi_3(\cos\theta_0, E, \lambda) & = & 
  \frac{d^2N}{dE d\lambda}
  \frac{1}{4\pi}
  \int_{T_{min}}^{T_{max}} \, dT \, T \frac{d^2N}{dT dx}                                    \label{eq:I3}
\end{eqnarray}

In this, it is assumed that the emission of photons from $\delta$ rays is isotropic.

\subsection{Light propagation}
\label{Light propagation}

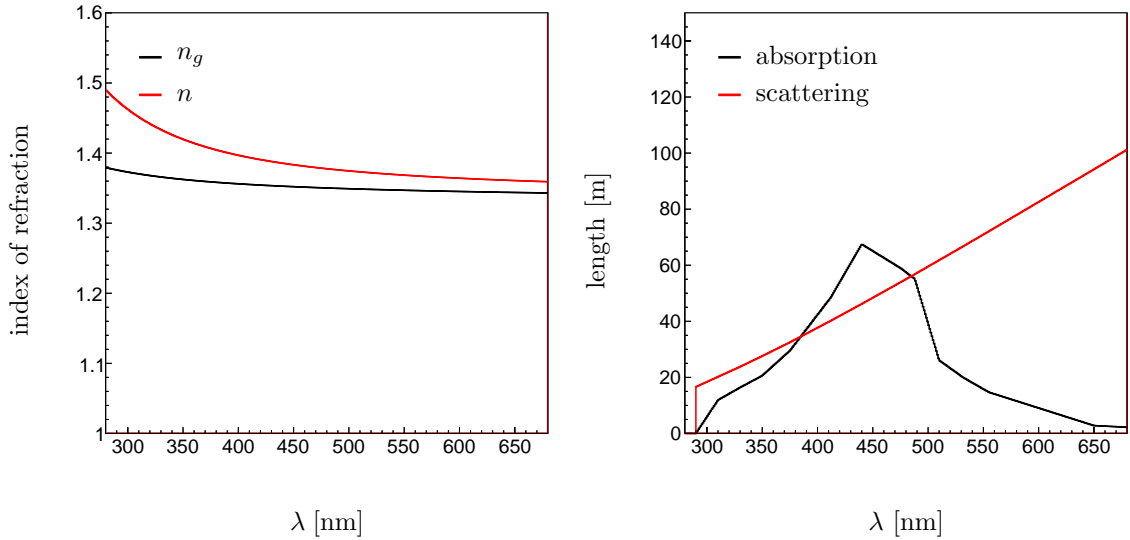
\begin{figure}[ht!]
\begin{center}
\begin{picture}(20,7)
\allinethickness{0.35mm}%
\put(4.6,-0.2){\makebox(0,0)[b]{\scalebox{0.4}{\includegraphics{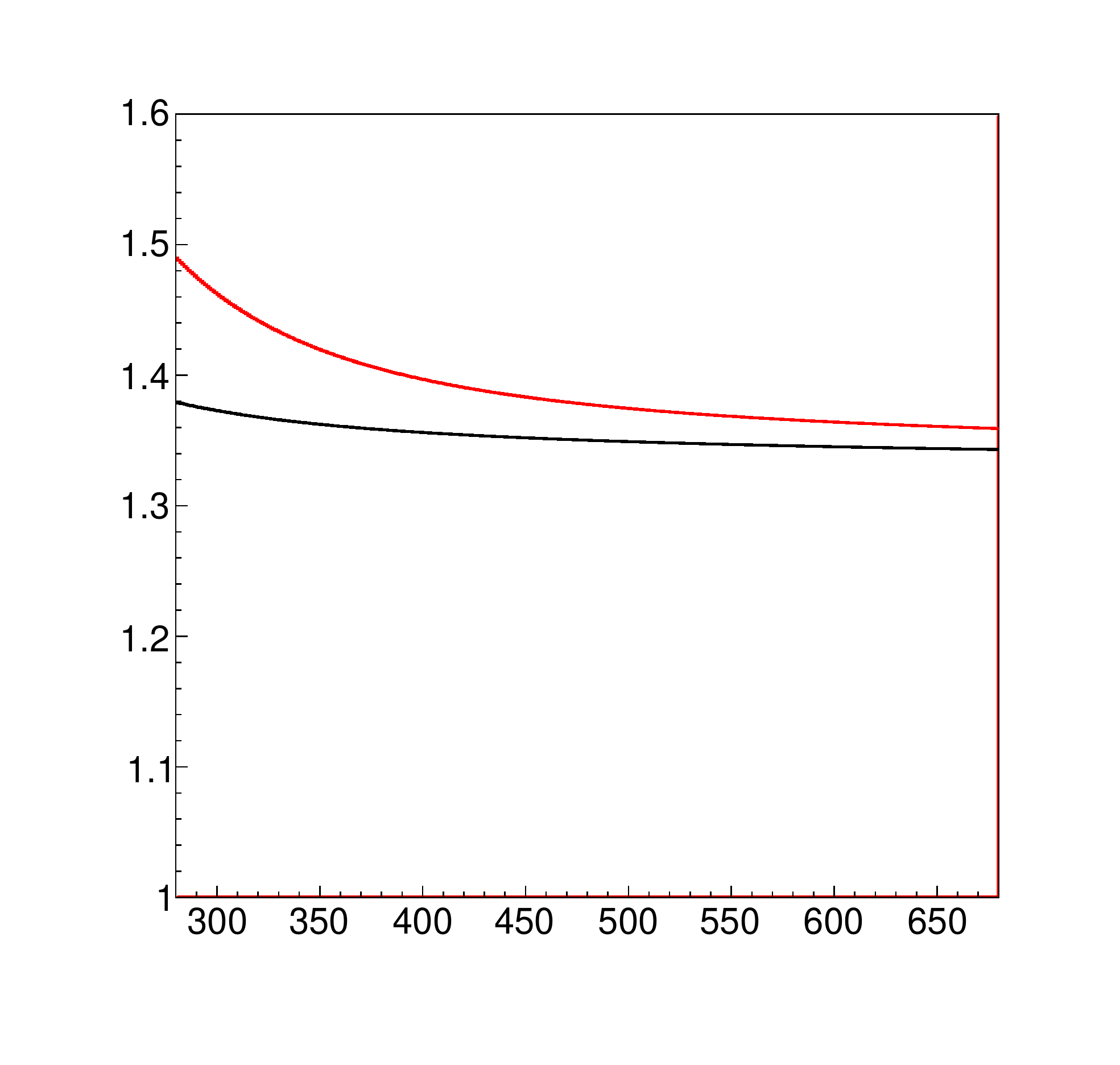}}}}%
\put(4.8,-0.2){\makebox(0,0){$\lambda\;[\mathrm{nm}]$}}%
\put(0.7, 3.7){\makebox(0,0){\rotatebox{90}{index of refraction}}}%
\put(2.3, 6.0){\color{black}\line(1,0){0.3}\color{black}}%
\put(2.8, 6.0){\makebox(0,0)[l]{$n_g$}}%
\put(2.3, 5.5){\color{red}\line(1,0){0.3}\color{black}}%
\put(2.8, 5.5){\makebox(0,0)[l]{$n$}}%
\put(12.3,-0.2){\makebox(0,0)[b]{\scalebox{0.4}{\includegraphics{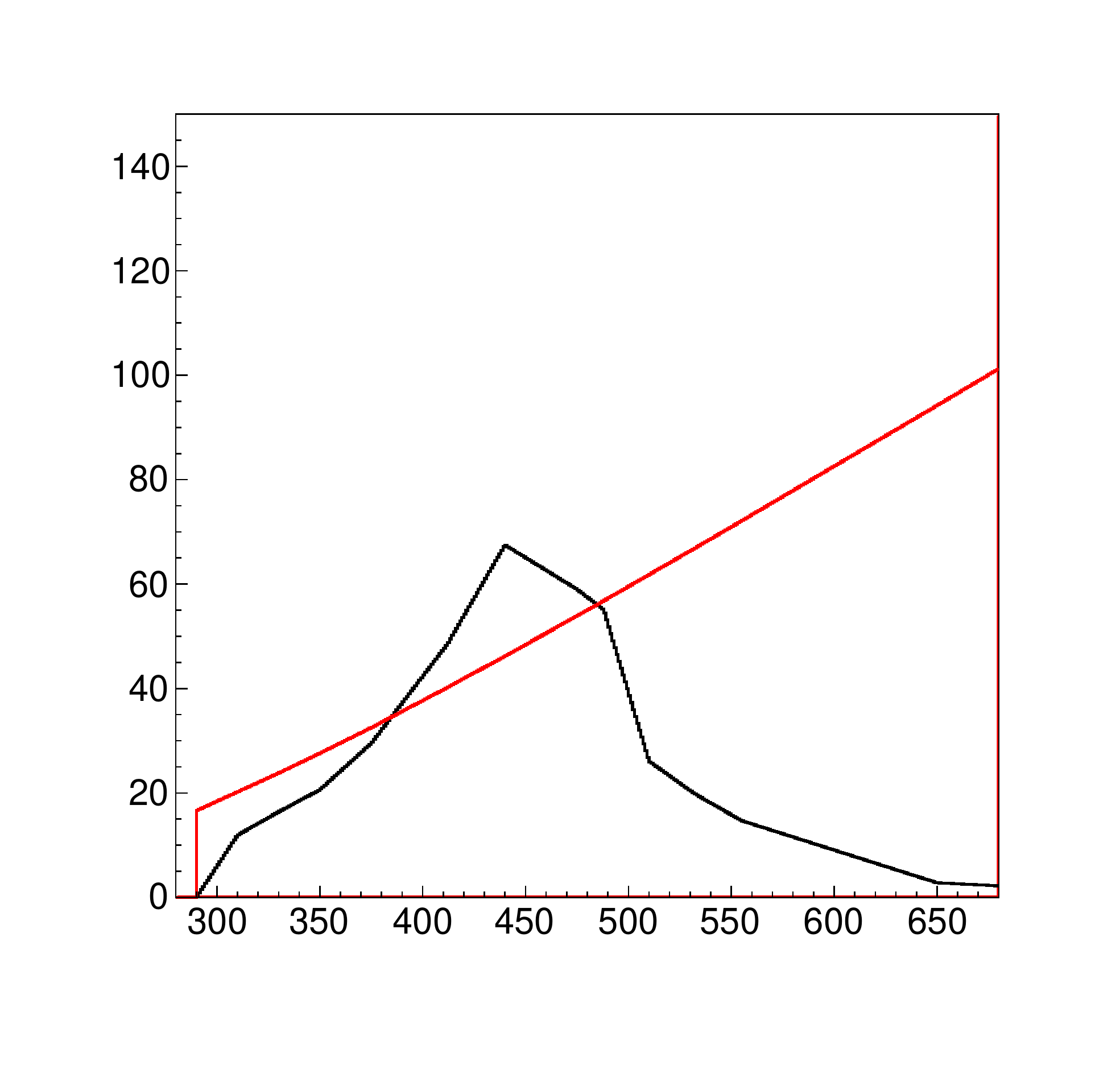}}}}%
\put(12.5,-0.2){\makebox(0,0){$\lambda ~[\mathrm{nm}]$}}%
\put( 8.4, 3.7){\makebox(0,0){\rotatebox{90}{length $[\mathrm{m}]$}}}%
\put(10.0, 6.0){\color{black}\line(1,0){0.3}\color{black}}%
\put(10.5, 6.0){\makebox(0,0)[l]{absorption}}%
\put(10.0, 5.5){\color{red}\line(1,0){0.3}\color{black}}%
\put(10.5, 5.5){\makebox(0,0)[l]{scattering}}%
\end{picture}
\end{center}
\caption{\label{f:light} 
Index of refraction and inverse of relative group velocity (left) and absorption length and scattering length (right) 
as a function of the wavelength of the light.
}
\end{figure}

In is commonly assumed that 
the phase velocity of light is related to the \v{C}erenkov angle and
the group velocity to the speed at which the light propagates through the medium.
The index of refraction, $n$, is defined as:

\begin{eqnarray}
  n  &  \equiv  &  c/v   \label{eq:index_of_refraction}
\end{eqnarray}

where 
$c$ refers to the speed of light (in vacuum) and
$v$ to the phase velocity of the light in the given medium.
The index of refraction, $n$, and inverse of the relative group velocity, $n_g = c/v_{g}$
are shown in figure \ref{f:light} as a function of the wavelength of the light, $\lambda$.

\subsection{Light absorption}
\label{Light absorption}

In general, light can be absorbed in the medium. 
This will affect the detected amount of light.
Absorption of light can be taken into account by introducing an extra term in the expression 
for the number of detectable photons, $\Phi$:

\begin{eqnarray}
  \Phi^{\prime} & = & \Phi \times e^{-d/\lambda_{abs}} \label{eq:absorption}
\end{eqnarray}

where 
$\lambda_{abs}$ refers to the absorption length and
$d$ to the distance traveled by the light.
The absorption length depends on the wavelength of the light.
A typical absorption length for deep-sea water as a function of the wavelength of the light is shown in figure \ref{f:light} \cite{ref:smith-baker}.

\subsection{Light scattering}
\label{Light scattering}

Various models exist that describe the effects of light scattering in deep-sea waters \cite{ref:bailey}.
Because the light scattering is rotation symmetric,
the scattering probability depends only on the space angle, $\theta_s$, 
which is defined as the angle between the direction of the light before and after the scattering.
Two commonly used light scattering models are presented in the following. 

\begin{itemize}
\item[-]
  The \texttt{f4} model is based on the so-called ``medsea'' parametrisation which is 
  a combination of two Henyey-Greenstein functions, each of which is defined as:

  \begin{eqnarray}
    f(a; \cos\theta_s) & = & \frac{1}{4\pi} \frac{1-a^2}{(1+a^2-2a\cos\theta_s)^{\frac{3}{2}}}
  \end{eqnarray}
  
  where $a$ is the average cosine of the scattering angle.
  This function is normalised to unity for the full solid angle.
  The parametrisation of the probability density function is then defined as:
  
  \begin{eqnarray}
    \frac{dP_s}{d\Omega_s} & = & p \times f(a_1; \cos\theta_s) ~ + ~ (1-p) \times f(a_2; \cos\theta_s) \label{eq:F4}
  \end{eqnarray}
  
  In the \texttt{f4} model, the values of $p$, $a_1$ and $a_2$ are respectively $1$, $0.77$ and $0$.
  
\item[-]
  The \texttt{p0.0075} model is based on a combination of 
  Rayleigh scattering and
  (geometric) scattering off large particles. 
  Rayleigh scattering is the elastic scattering of light by particles that are typically much smaller 
  than the wavelength of the light.
  The corresponding cross section can be expressed as \cite{ref:rayleigh}:

  \begin{eqnarray}
    \frac{d\sigma}{d\Omega_s}  & = & \frac{\pi^4}{8} \left(\frac{n^2 - 1}{n^2 + 2}\right)^2 \frac{d^6}{\lambda^4} (1 + \cos^2\theta_s)   \label{eq:rayleigh}
  \end{eqnarray}
  
  where 
  $n$ is the index of refraction of the medium,
  $d$ the diameter of the particle and
  $\lambda$ the wavelength of the light.
  In the \texttt{p0.0075} model, a slightly different parametrisation for the angular distribution is assumed
  to take into account the anisotropy of the water molecules:
  
  \begin{eqnarray}
    g(a, b; \cos\theta_s) & = & a \: (1 + b \cos^2\theta_s)
  \end{eqnarray}
  
  where $a = 0.06225$ and $b = 0.835$.
  The (geometric) scattering off large particles is well described by Mie's solution of the Maxwell equations.
  In the \texttt{p0.0075} model, the distribution for the scattering off large particles is 
  obtained from measurements {\em in situ}.
  The average cosine of the scattering angle has been measured and is found to be $0.924$.
  A Henyey-Greenstein function is used which leads to the same average cosine.
  The parametrisation of the probability density function is then defined as:
  
  \begin{eqnarray}
    \frac{dP_s}{d\Omega_s} & = & p \times g(\cos\theta_s) ~ + ~ (1-p) \times f(a; \cos\theta_s) \label{eq:p0.0075}
  \end{eqnarray}
  
  where $a = 0.924$.
  In the \texttt{p0.0075} model, the relative contribution of the Rayleigh function is set to $p = 0.17$.
  
\end{itemize}

\begin{figure}[ht!]
\begin{center}
\begin{picture}(8,7)
\allinethickness{0.35mm}%
\put(4.2,-0.1){\makebox(0,0)[b]{\scalebox{0.4}{\includegraphics{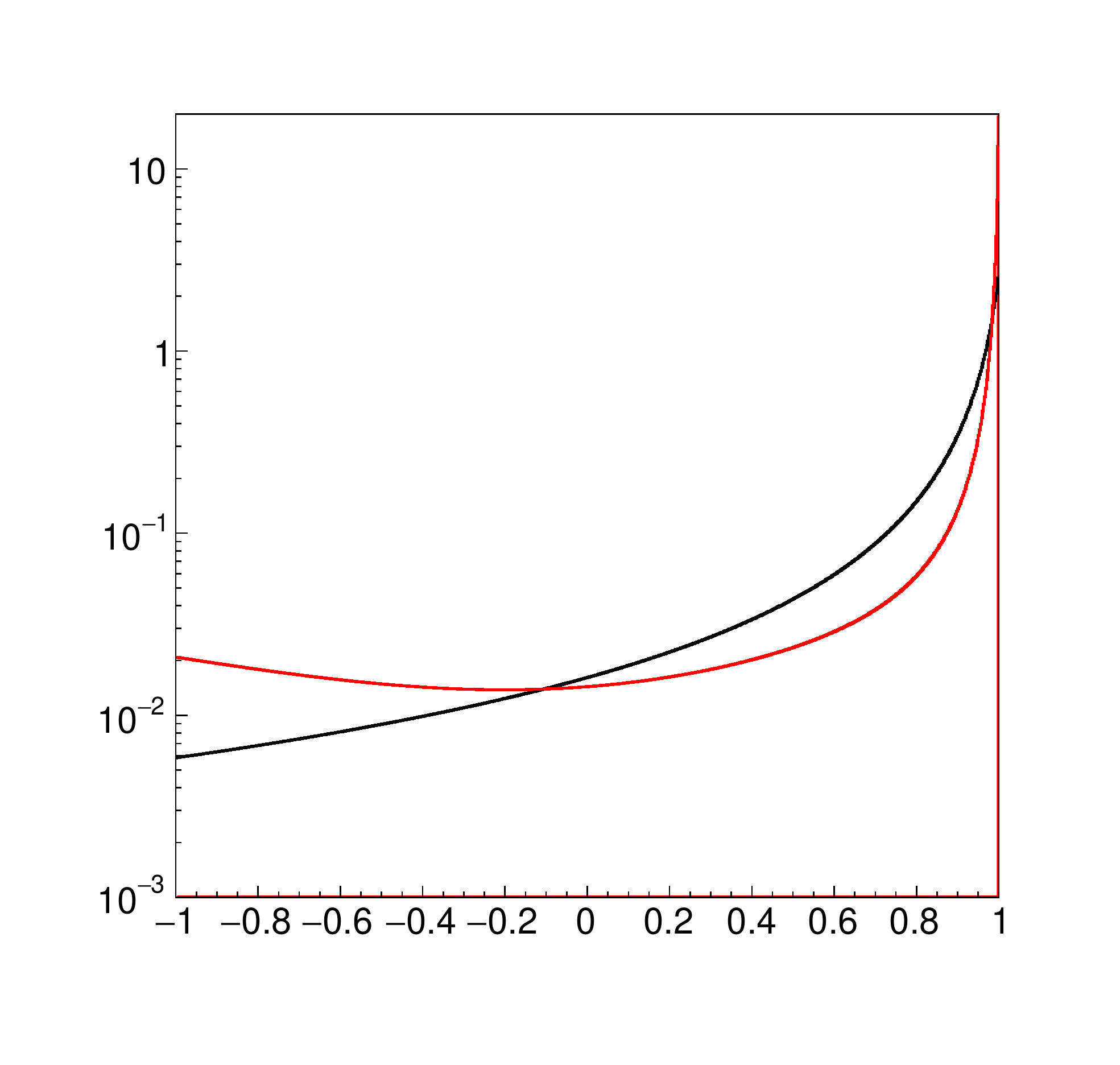}}}}%
\put(4.2,-0.2){\makebox(0,0){$\cos\theta_s$}}%
\put(0.0, 3.7){\makebox(0,0){\rotatebox{90}{$\frac{\textstyle dP}{\textstyle d\Omega_s}$}}}%
\put(4.5, 6.0){\color{black}\line(1,0){0.3}\color{black}}%
\put(5.0, 6.0){\makebox(0,0)[l]{\texttt{f4}}}%
\put(4.5, 5.5){\color{red}\line(1,0){0.3}\color{black}}%
\put(5.0, 5.5){\makebox(0,0)[l]{\texttt{p0.0075}}}%
\end{picture}
\end{center}
\caption{\label{f:Ps} 
Parametrisations of the angular distributions of light scattering of two commonly used models.
}
\end{figure}
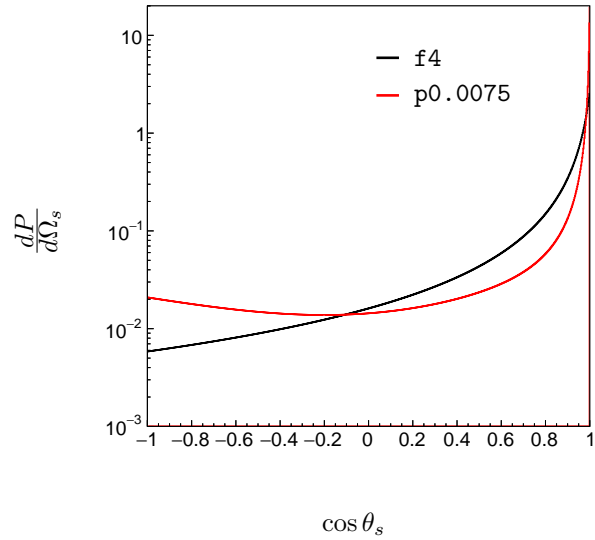

The distributions of the scattering angles of the \texttt{f4} and \texttt{p0.0075} models 
are shown in figure \ref{f:Ps}. 
The number of light scatterings per unit track length can be expressed as:

\begin{eqnarray}
  -\frac{dN}{dx} & = & \frac{N}{\lambda_{s}} \label{eq:scattering-length}
\end{eqnarray}

where 
$N$ refers to the number of photons and
$\lambda_{s}$ to the scattering length.
Due to the scattering of light, 
the path of the photon is not uniquely defined (see below).
The scattering will also affect the amount of direct (i.e.\ not scattered) light that can be detected.
This can be taken into account by introducing an extra term in the expression 
for the number of detectable photons, $\Phi$:

\begin{eqnarray*}
  \Phi^{\prime} & = & \Phi \times e^{-d/\lambda_{s}}
\end{eqnarray*}

where 
$d$ refers to the distance traveled by the light.
For indirect light, an effective attenuation length is used which is applied to the calculated intensity of single-scattered light:

\begin{eqnarray}
  \frac{1}{\lambda_{att}} & = & \frac{1}{\lambda_{abs}} + \frac{w}{\lambda_{s}}  \label{eq:attenuation}
\end{eqnarray}

For direct light, $w = 1$.
For indirect light, $w$ is usually defined as $1 - \left<\cos\theta_s\right>$.
Here, the weight is defined as:
\begin{eqnarray*}
  w(\cos\theta) & = & \int_{-1}^{\cos\theta_s} \frac{dP_s}{d\Omega_s} \: 2\pi \: d\cos\theta
\end{eqnarray*}

As a result, light which scattered at a small (large) angle is more (less) attenuated compared
to the usual definition.
\newline

For the application of either model in the Monte Carlo simulation (see below), 
it is assumed that the dependence of the scattering length on 
the wavelength of the light is identical for the different contributions to the light scattering.
The assumed common scattering length is shown in figure \ref{f:light}.
As can be seen from figure \ref{f:light}, 
the scattering length increases somewhat between linearly and quadratically with the wavelength of the light.
For scattering off large particles (``Mie scattering''),  
the scattering length does not depend strongly on the wavelength of the light.
For Rayleigh scattering,
the scattering length should increase with the fourth power of the wavelength of the light (see equation \ref{eq:rayleigh}).
This apparent discrepancy has not been resolved.

\subsection{Light detection}
\label{Light detection}

\begin{figure}[ht!]
\begin{center}
\begin{picture}(20,7)
\allinethickness{0.35mm}%
\put(4.5,-0.2){\makebox(0,0)[b]{\scalebox{0.4}{\includegraphics{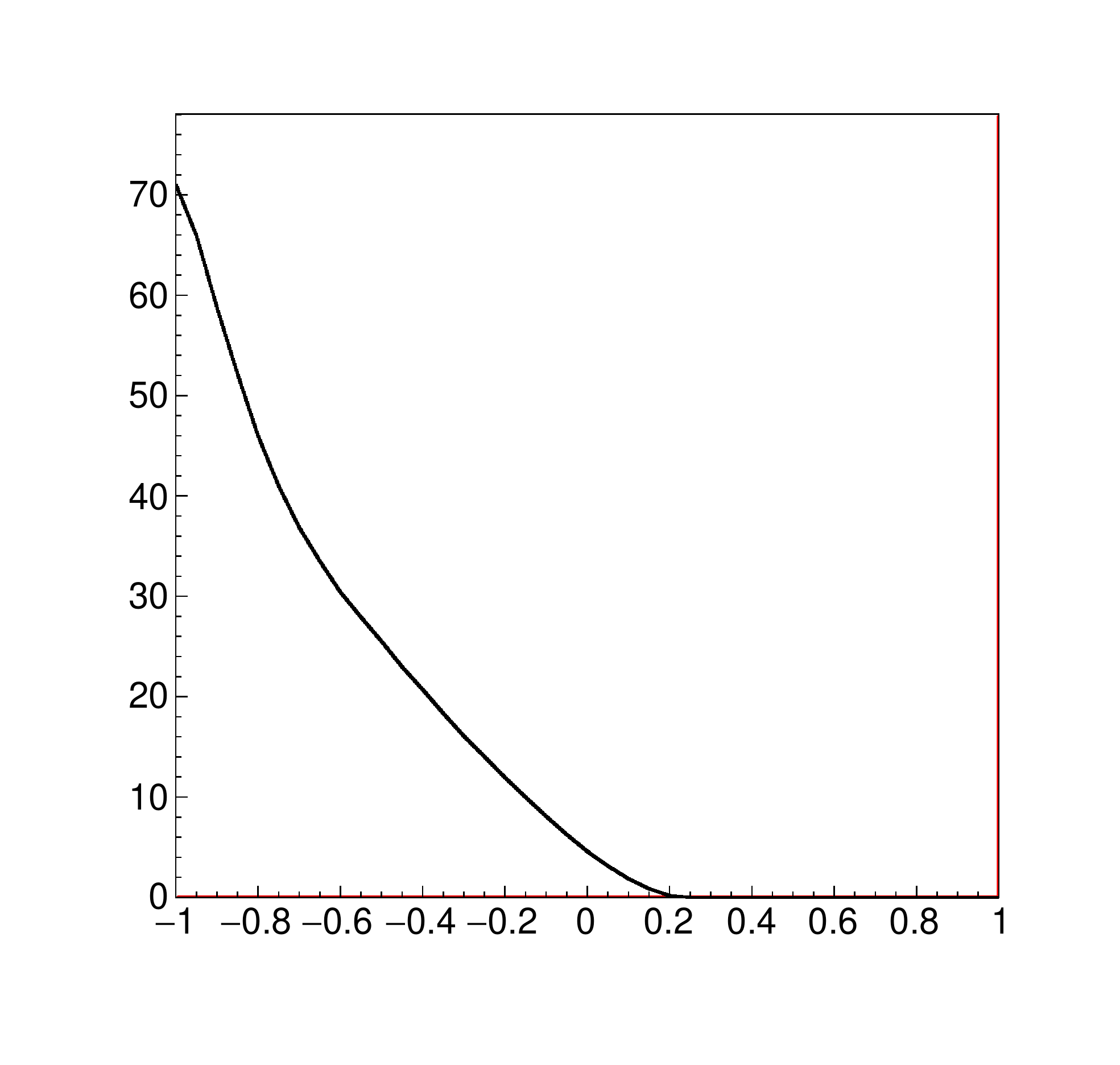}}}}%
\put(4.8,-0.2){\makebox(0,0){$\cos\theta_\oslash$}}%
\put(0.7, 3.7){\makebox(0,0){\rotatebox{90}{$\varepsilon$}}}%
\put(4.3, 6.0){\color{black}\line(1,0){0.3}\color{black}}%
\put(4.8, 6.0){\makebox(0,0)[l]{acceptance}}%
\put(12.2,-0.2){\makebox(0,0)[b]{\scalebox{0.4}{\includegraphics{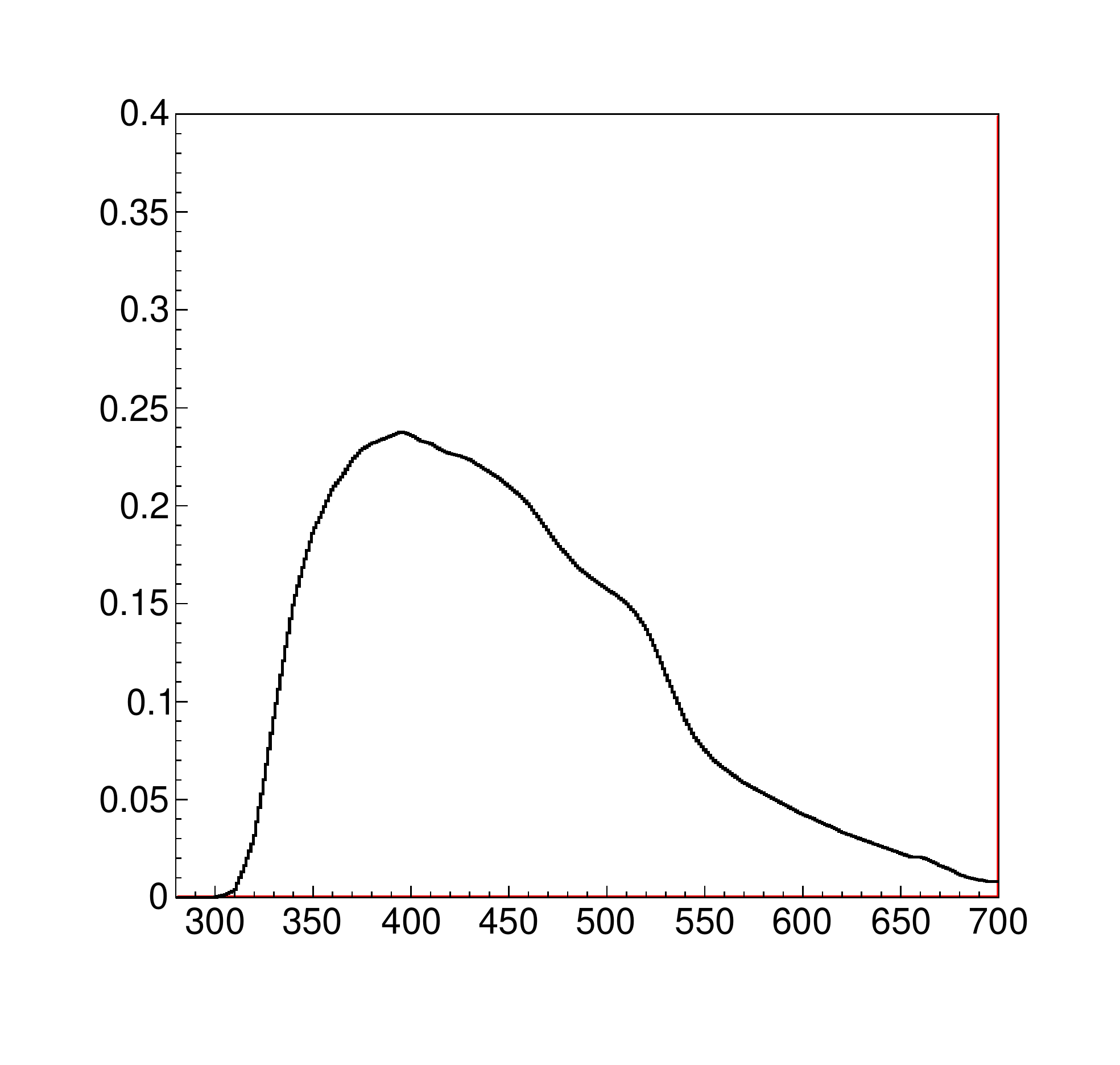}}}}%
\put(12.5,-0.2){\makebox(0,0){$\lambda ~[\mathrm{nm}]$}}%
\put( 8.4, 3.7){\makebox(0,0){\rotatebox{90}{QE}}}%
\put(12.0, 6.0){\color{black}\line(1,0){0.3}\color{black}}%
\put(12.5, 6.0){\makebox(0,0)[l]{QE}}%
\end{picture}
\end{center}
\caption{\label{f:pmt} 
Parametrisations of the angular acceptance of the PMT as a function of the cosine of the angle of incidence (left) and 
the quantum efficiency of the PMT as a function of the wavelength of the light (right).
}
\end{figure}
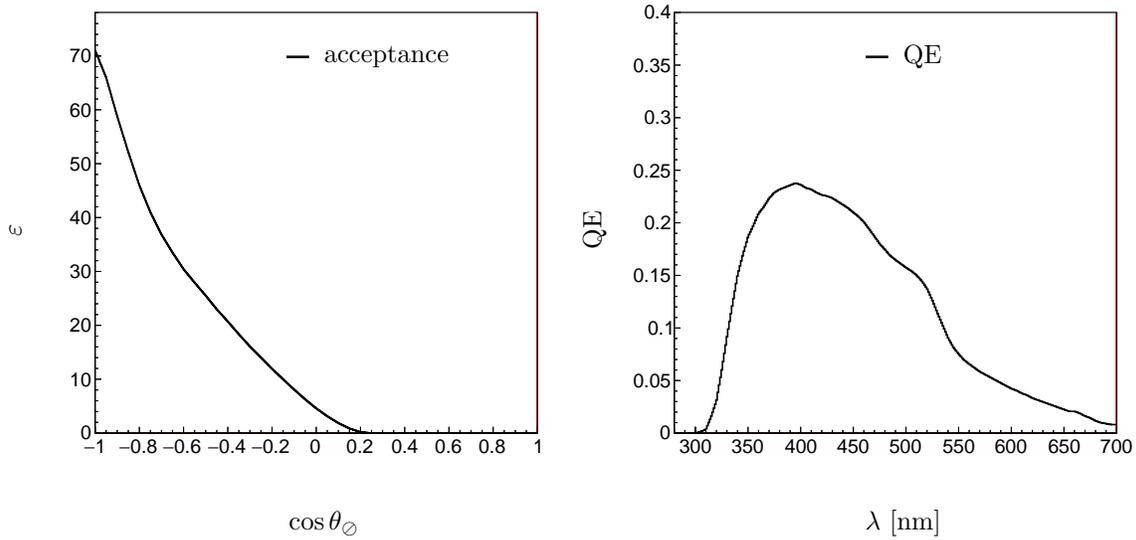

The light is detected using a PMT.
The acceptance of a PMT depends primarily on 
the wavelength, $\lambda$, and 
the angle of incidence, $\theta_\oslash$, 
of the photon.
This angle is defined as the angle between 
the direction of the photon and
the axis of the PMT (see figure \ref{f:event}).
Here, it is implicitly assumed that the acceptance of the PMT is independent of 
the azimuth angle and 
the impact point of the photo-cathode area.
The angular acceptance and the quantum efficiency of a typical 3'' PMT are shown in figure \ref{f:pmt}.
The quantum efficiency, QE, shown in figure \ref{f:pmt} includes the collection efficiency and the transparency of the glass sphere.

The cosine of the angle of incidence can be determined from 
the direction of the photon and the orientation of the PMT, i.e:

\begin{eqnarray}
  \cos\theta_\oslash & = & 
  \left( 
  \begin{array}{c}
    \sin\theta_{\wp}\cos\phi_{\wp} \\
    \sin\theta_{\wp}\sin\phi_{\wp} \\
    \cos\theta_{\wp}
  \end{array}
  \right)
  \cdot 
  \left( 
  \begin{array}{c}
    \sin\theta_{1}\cos\phi_{1} \\
    \sin\theta_{1}\sin\phi_{1} \\
    \cos\theta_{1}
  \end{array}
  \right)
  \label{eq:PMT}
\end{eqnarray}

where 
$\theta_{\wp}$ and $\phi_{\wp}$ correspond to the zenith and azimuth angle of the orientation of the PMT and 
$\theta_{1}$   and $\phi_{1}$   to the zenith and azimuth angle of the direction of the photon.
In the absence of light scattering, equation \ref{eq:PMT} reduces to:

\begin{eqnarray}
  \cos\theta_\oslash  & = & \sin\theta_\wp \cos\phi_\wp \sin\theta_0 +  \cos\theta_\wp \cos\theta_0 
\end{eqnarray}

The solid angle of a PMT, $d\Omega$, is defined as:

\begin{eqnarray}
    d\Omega & \equiv & \frac{A}{\:d^2}  \label{eq:omega}
\end{eqnarray}

where 
$A$ refers to the photo-cathode area and
$d$ to the distance traveled by the light.

\section{Probability Density Functions}
\label{Probability Density Functions}

In this section, 
the probability density functions (PDFs)
of the arrival times of photons due to various sources of light are presented.
In the absence of light scattering, 
the expected arrival time is completely determined by 
the position $z$ or the angle $\theta_0$ of the emitted photon 
and the velocity of light (see figure \ref{f:event}).

\begin{eqnarray}
  ct       & = & z + n_g\sqrt{z^2+R^2}                                              \label{eq:time-direct-z}  \\
  \mbox{}  & = & -\frac{R}{\tan\theta_0} + n_g\frac{R}{\sin\theta_0}                \label{eq:time-direct-ct} 
\end{eqnarray}

In the absence of light dispersion, 
the earliest possible arrival time, $t_0$, is sharply defined. 
This time corresponds by definition to the shortest optical path.
Due to light dispersion, 
the shortest optical path gets smeared.
The dependence of the arrival time on the wavelength of the light, $\lambda$, should then be considered.
If the light is emitted from a fixed position (e.g.\ in the case of light from a shower),
the dependence of the arrival time on the wavelength of the light can be formulated as:

\begin{eqnarray}
  \frac{\partial ct}{\partial \lambda} & = & d \, \frac{\partial n_g}{\partial \lambda} \label{eq:dt'/dlambda}
\end{eqnarray}

where $d$ refers to the distance traveled by the light.
When the wavelength dependence of the \v{C}erenkov angle should be taken into account as well, 
the derivative of the arrival time as a function of $\lambda$ becomes:

\begin{eqnarray}
  \frac{\partial ct}{\partial \lambda} & = & R \; 
  \left(
  \frac{1}{\sin\theta_0} \frac{dn_g}{d\lambda} + 
  \frac{n - n_g}{\tan^3\theta_0}\frac{dn}{d\lambda} 
  \right)                                                                           \label{eq:dt/dlambda}
\end{eqnarray}

For the light due to the energy loss of a muon, 
the dependence of the arrival time on the position $z$ should be considered.
The derivative of the arrival time as a function of $z$ can be formulated as:

\begin{eqnarray}
  \frac{\partial ct}{\partial z} & = & 1 - n_g\cos\theta_0                          \label{eq:dt/dz}
\end{eqnarray}

As can be seen from equation \ref{eq:dt/dz}, 
the derivative of the time is zero at $\theta_0 = \theta_C$.
The second derivative of the arrival time as a function of $z$ can be formulated as:

\begin{eqnarray}
  \frac{\partial^2 ct}{(\partial z)^2} & = & n_g \frac{\sin^3\theta_0}{R}           \label{eq:d2t/dz2}      
\end{eqnarray}

As can be seen from equation \ref{eq:d2t/dz2}, the second derivative in strictly positive.
This shows that indeed there is an arrival time, $t_0$, that corresponds to the shortest optical path.
Assuming that the effect of dispersion of light is small, 
this implies that the distribution of the arrival time of any light will exhibit a leading edge at $t = t_0$.
\newline

Due to scattering of light, 
the path of the photon is not uniquely defined.
Assuming a single scattering of the light, the arrival time can be expressed as:

\begin{eqnarray}
  ct       & = & z + n_g (u + v)                                                    \label{eq:time-indirect}
\end{eqnarray}

where 
$u \equiv \left|\bar{u}\right|$ and 
$v \equiv \left|\bar{v}\right|$ 
refer to the distances traveled by the photon 
before and after the scattering, respectively (see figure \ref{f:scattering}). 
The various paths of the photon are constrained by the following geometrical condition:

\begin{eqnarray}
  \left( 
  \begin{array}{c}
    R \\
    0 \\
    0 
  \end{array}
  \right)  
  & = &
  \left( 
  \begin{array}{c}
    0 \\
    0 \\
    z 
  \end{array}
  \right)  
  + u
  \left( 
  \begin{array}{c}
    \sin\theta_0 \cos\phi_0 \\
    \sin\theta_0 \sin\phi_0 \\
    \cos\theta_0            \\
  \end{array}
  \right)  
  + v
  \left( 
  \begin{array}{c}
    \sin\theta_1 \cos\phi_1 \\
    \sin\theta_1 \sin\phi_1 \\
    \cos\theta_1            \\
  \end{array}
  \right)                   \label{eq:position-indirect}
\end{eqnarray}

The term on the left hand side corresponds to the position of the PMT and
the terms on the right hand side correspond to 
the point of emission of the photon along the muon trajectory,
the path upstream   of the scattering ($\bar{u}$) and
the path downstream of the scattering ($\bar{v}$),
respectively.
\newline

In order to evaluate the total PDF, 
one should integrate over the full range of directions of the emitted photons  
provided that equations \ref{eq:time-indirect} and \ref{eq:position-indirect} are satisfied.
For the PDF of light from energy loss processes, one should also integrate over all $z-$positions. 
In general, this requires a summation using discrete values for $\cos\theta_0$, $\phi_0$ and $z$.
For each set of values, equations \ref{eq:time-indirect} and \ref{eq:position-indirect} 
should then be solved explicitly.
This is possible because there are 4 equations and 4 un-knowns, namely $u$, $v$, $\theta_1$ and $\phi_1$.
The solution to this problem can be summarised as:

\begin{eqnarray}
  d  & = & \frac{ct - z}{n_g}                                                        \label{eq:d} \\
  u  & = & \frac{R^2 + z^2 - d^2}{2R \sin\theta_0 \cos\phi_0 - 2z \cos\theta_0 - 2d} \label{eq:u} \\*[2mm]
  v  & = & d - u                                                                     \label{eq:v}
\end{eqnarray}

The cosine of the scattering angle can be obtained directly from
equation \ref{eq:position-indirect} by multiplying the left hand side and the right hand side with $\hat{u}$ 
(see equation \ref{eq:scattering-angle} for the definition of the scattering angle), i.e:

\begin{eqnarray}
  \cos\theta_s = \frac{R\sin\theta_0\cos\phi_0 - z\cos\theta_0 - u}{v}
\end{eqnarray}

Due to the scattering of the light,
the solid angle of the PMT should now be evaluated from 
the point where the scattering took place instead of
the point where the photon has been emitted.
In general, the number of scattered photons is equal to the product of 
some photon flux, $\Phi$, 
the volume $V = A \, du$ and 
the inverse of the scattering length.
For a \v{C}erenkov light cone, 
$\Phi = \Phi_0(u\sin\theta_C)$ and $A = u\sin\theta_C \, d\phi_0 \, sin^2\theta_C \, dz$.
For light from a shower, $\Phi = \Phi_1(\cos\theta) \, u^{-2}$ and $A = u^2 d\cos\theta_0d\phi_0$.
In both cases, the number of scattered photons does not depend on the distance $u$ \cite{ref:kooijman}.
The solid angle is thus:

\begin{eqnarray}
    d\Omega & = & \frac{A}{\:v^2} \label{eq:solid-angle}
\end{eqnarray}

It should be noted that for small $v$, the solid angle is of course limited to $2\pi$.
\newline

The direction of the photon after the scattering 
is needed to determine the angle of incidence on the PMT.
The unit direction vector $\hat{v}$ is completely determined by the solution above, 
and is given here for completeness.

\begin{eqnarray}
  \hat{v} & = &
  \frac{1}{v}
  \left( 
  \begin{array}{r@{\;}c@{\;}c@{\,}l@{\,}l}
    R  & - & u & \sin\theta_0 & \cos\phi_0 \\
       & - & u & \sin\theta_0 & \sin\phi_0 \\
    -z & - & u & \cos\theta_0 &            
  \end{array}
  \right)                                 \label{eq:vhat}
\end{eqnarray}

In the case of light scattering, 
the dependence of the arrival time on the length $u$ should be considered 
($\cos\theta_0$, $\phi_0$, and $z$ have been fixed).
The derivative of the arrival time as a function of the length $u$ 
can be formulated as (see equation \ref{eq:time-indirect}):

\begin{eqnarray}
  \frac{\partial ct}{\partial u} & = & n_g \left(1 + \frac{\partial v}{\partial u}\right) \label{eq:dt/du}
\end{eqnarray}

where

\begin{eqnarray}
  \frac{\partial v}{\partial u}  & = & -\cos\theta_s  \label{eq:dv-du}
\end{eqnarray}

The derivation is given in Appendix \ref{A:derivative}.
It is obvious but worth noting that for very small scattering angles, 
the arrival time does not depend on the length $u$.
Or, in other words, the arrival time of the light no longer depends 
on the location of a scattering point along a line that is almost straight. 
As a result, the probability density function, $P_s$, for the scattering of light is 
weighed with a function that exhibits a pole at $\cos\theta_s = 1$.
\newline

For the light from a shower, 
the corresponding PDF should be convoluted with the shower profile which 
depends on the energy of the shower (see equation \ref{eq:dP/dz}).

\subsection{Direct light from a muon}
\label{Direct light from a muon}

For direct light from the muon, the zenith angle at which the photons are emitted 
can be considered fixed ($\theta_0 = \theta_C$).
The distribution of the arrival times of the photons is then mainly determined 
by the dispersion of light in the medium.
The probability density function for the distribution of the arrival times can then be expressed as:

\begin{eqnarray}
  \frac{d\mathcal{P}}{dt} & \!\!=\!\! & 
  \Phi_0(R,\lambda) \, A                                                            \;
  \left(\frac{\partial t}{\partial \lambda}\right)^{-1}                             \:
  \varepsilon(\cos\theta_\oslash)                                                   \;
  QE(\lambda)                                                                       \;
  e^{-d/\lambda_{abs}} \: e^{-d/\lambda_{s}}                                        \label{eq:f0}
\end{eqnarray}

where 
$\Phi_0()$ is the detectable photon flux per unit wavelength as a function of $R$ (equation \ref{eq:I0}), 
$A$ the photo-cathode area of the PMT, 
$\varepsilon$ the angular acceptance of the PMT as a function of the angle of incidence of the photon, and
$QE$ the quantum efficiency of the PMT as a function of the wavelength.
The wavelength, $\lambda$, is constraint by equation \ref{eq:time-direct-z}.
The derivative of the time is given by equation \ref{eq:dt/dlambda}.
The distance traveled by the photons and the \v{C}erenkov angle are related as $d = \sqrt{R^2 + z^2} = R/\sin\theta_C$.
It is interesting to note that due to the $R$ dependence of the time derivative,
the number of photons detected in a small time window decreases with the square of the distance $R$ (see equation \ref{eq:dt/dlambda}).
The integrated signal decreases 
--as expected from the \v{C}erenkov cone hypothesis-- 
linearly with the distance $R$.

\subsection{Direct light from a shower}
\label{Direct light from a shower}

In general, the light from a shower is emitted at all angles.
As a consequence, the amount of detected light is proportional to the solid angle of the PMT.
The probability density function per unit energy for the distribution of the arrival times can then be expressed as:

\begin{eqnarray}
  \frac{d^2\mathcal{P}}{dE dt} & \!\!=\!\! & 
  \Phi_1(\cos\theta_0, \lambda)                                                     \;
  \left(\frac{\partial t}{\partial \lambda}\right)^{-1}                             \:
  \varepsilon(\cos\theta_\oslash)                                                   \;
  QE(\lambda)                                                                       \;
  e^{-d/\lambda_{abs}} \: e^{-d/\lambda_{s}}                                        \;
  d\Omega                                                                           \label{eq:f1}
\end{eqnarray}

where 
$\Phi_1()$ refers to the detectable photon flux per unit energy as a function of $\cos\theta_0$ (equation \ref{eq:I1}) and
$d\Omega$ to the solid angle of the PMT (equation \ref{eq:omega}).
The wavelength, $\lambda$, is constraint by equation \ref{eq:time-direct-z}.
The derivative of the time is given by equation \ref{eq:dt'/dlambda}.

\subsection{Direct light due to the energy loss of a muon}
\label{Direct light due to the energy loss of a muon}

For an arrival time that is later than the shortest optical path, 
equation \ref{eq:time-direct-z} has two solutions, namely:

\begin{eqnarray}
  z_{1,2} & = & \frac{-b \pm \sqrt{b^2 - 4ac}}{2a}  \label{eq:z12}
\end{eqnarray}

where
\begin{eqnarray*}
  a & = & n_g^2 - 1          \\
  b & = & 2ct                \\
  c & = & (Rn_g)^2 - (ct)^2   
\end{eqnarray*}

The corresponding distance traveled by the photon and the angle at which the light is emitted 
can be formulated respectively as:

\begin{eqnarray*}
  d~~~         & = & \sqrt{R^2 + z^2} \\
  \cos\theta_0 & = & -z/d     
\end{eqnarray*}

For $\cos\theta_0 = 1/n_g$, the derivative of the arrival time as a function of $z$ becomes zero (see equation \ref{eq:dt/dz}).
The second derivative of the arrival time as a function of $z$ should therefore also be considered (equation \ref{eq:d2t/dz2}).
To second order approximation, the dependence of the arrival time as a function of $z$ can then be formulated as:

\begin{eqnarray}
  \frac{d t}{d z} & = & \frac{\partial t}{\partial z} ~ + ~ \frac{1}{2} \frac{D}{\sin\theta_0} \frac{\partial^2 t}{(\partial z)^2} \label{eq:Dt/Dz}
\end{eqnarray}

where $D$ is the diameter of the photo-cathode area ($D = 2\sqrt{A/\pi}$).
The probability density function for the distribution of the arrival times can then be expressed as:

\begin{eqnarray}
  \frac{d\mathcal{P}}{dt} & \!\!=\!\! & 
  \int d\lambda                                                                     \:
  \sum_{z=z_1,z_2}                                                                  \:
  \left(\frac{d t}{d z}\right)^{-1}                                                 \:
  \Phi_{2,3}(\cos\theta_0, E, \lambda) \, d\Omega                                   \;
  \varepsilon(\cos\theta_\oslash)                                                   \;
  QE(\lambda)                                                                       \;
  e^{-d/\lambda_{abs}} \: e^{-d/\lambda_{s}}                                        \label{eq:f3}
\end{eqnarray}

where 
$\Phi_{2,3}()$ refers to the detectable photon intensity 
per unit wavelength,
per unit track length and 
per unit solid angle 
as a function of 
$\cos\theta_0$ and 
$E$ (equation \ref{eq:I2}, \ref{eq:I3}, respectively), and
$d\Omega$ to the solid angle of the PMT (equation \ref{eq:omega}).
The derivative of the time is given by equation \ref{eq:Dt/Dz}.

\subsection{Indirect light from a muon}
\label{Indirect light from a muon}

For indirect light from the muon, one has to integrate over the full range of 
azimuth angles and positions of the photon emission points.
The probability density function for the distribution of the arrival times can then be expressed as:

\begin{eqnarray}
  \frac{d\mathcal{P}}{dt} & \!\!=\!\! & 
  \int\!\!\!\int\!\!\!\int d\lambda \, dz \, d\phi_0                                \:
  \frac{1}{2\pi}\frac{d^2N}{dx d\lambda}                                                       \,
  \frac{1}{\lambda_{s}}                                                             \,
  \left(\frac{\partial t}{\partial u}\right)^{-1}\!                                 \,
  \varepsilon(\cos\theta_\oslash)                                                   \;
  QE(\lambda)                                                                       \;
  e^{-d/\lambda_{att}}                                                              \;
  \frac{dP_s}{d\Omega_s}                                                            \,
  d\Omega                                                                           \label{eq:f4}
\end{eqnarray}

where
$d\Omega$ refers to the solid angle of the PMT (equation \ref{eq:solid-angle}).
The factor $2\pi$ takes into account the re-normalisation of 
the number of detectable photons per unit track length (equation \ref{eq:I0})
due to the explicit integration over $\phi_0$.
The term $1/\lambda_{s}$ corresponds to the probability for the scattering of the light per unit length.
The derivative of the time is given by equation \ref{eq:dt/du}.
The distance traveled by the photons is defined as $d = u + v$ (see above).
\newline

The lower and upper limit for the integral of $z$ can be determined 
using equation \ref{eq:z12}.
These values correspond to the case that the photon would scatter immediately 
after the point of emission and continues to travel to the PMT in a straight line.
Apart from the orientation of the PMT, equation \ref{eq:f4} exhibits 
a mirror symmetry in the $x-z$ plane.
One can thus integrate $\phi_0$ between $0$ and $\pi$ and 
evaluate for each $\phi_0$ the angle of incidence of the PMT twice 
using equation \ref{eq:PMT}, 
i.e.\ substituting $\phi_0$ and $-\phi_0$ in equation \ref{eq:vhat}.

\subsection{Indirect light from a shower}
\label{Indirect light from a shower}

For indirect light from a shower, one has to integrate over the full range of 
zenith and azimuth angles of the photon emission profile.
The probability density function per unit energy for the distribution of the arrival times can then be expressed as:

\begin{eqnarray}
  \frac{d^2\mathcal{P}}{dE dt}  & = & 
  \int\!\!\!\int\!\!\!\int d\lambda \, d\phi_0 \, d\!\cos\theta_0                   \:
  \Phi_1(\cos\theta_0, \lambda)                                                     \,
  \frac{1}{\lambda_{s}}                                                             \,
  \left(\frac{\partial t}{\partial u}\right)^{-1}\!                                 \,
  \varepsilon(\cos\theta_\oslash)                                                   \;
  QE(\lambda)                                                                       \;
  e^{-d/\lambda_{att}}                                                              \;
  \frac{dP_s}{d\Omega_s}                                                            \,
  d\Omega                                                                           \label{eq:f5}
\end{eqnarray}

where
$\Phi_1()$ refers to the detectable photon flux per unit energy as a function of $\cos\theta_0$ (equation \ref{eq:I1}) and
and $d\Omega$ to the solid angle of the PMT (equation \ref{eq:solid-angle}).
The term $1/\lambda_{s}$ corresponds to the probability for the scattering of the light per unit length. 
The derivative of the time is given by equation \ref{eq:dt/du}.
The distance traveled by the photons is defined as $d = u + v$ (see above).

\subsection{Indirect light due to the energy loss of a muon}
\label{Indirect light due to the energy loss of a muon}

For indirect light due to the energy loss of a muon, one has to integrate over the full range of 
zenith and azimuth angles and positions of the photon emission points.
The probability density function for the distribution of the arrival times can then be expressed as:

\begin{eqnarray}
  \frac{d\mathcal{P}}{dt}  & \!\!=\!\! & 
  \int\!\!\!\int\!\!\!\int\!\!\!\int d\lambda \, dz \, d\phi_0 \, d\!\cos\theta_0   \:
  \Phi_{2,3}(\cos\theta_0, E, \lambda)                                              \,
  \frac{1}{\lambda_{s}}                                                             \,
  \left(\frac{\partial t}{\partial u}\right)^{-1}\!                                 \,
  \varepsilon(\cos\theta_\oslash)                                                   \;
  QE(\lambda)                                                                       \;
  e^{-d/\lambda_{att}}                                                              \;
  \frac{dP_s}{d\Omega_s}                                                            \,
  d\Omega                                                                           \label{eq:f6}
\end{eqnarray}

where
$\Phi_{2,3}()$ refers to the number of detectable photons 
per unit wavelength,
per unit track length and 
per unit solid angle 
as a function of 
$\cos\theta_0$ and 
$E$ (equation \ref{eq:I2} and \ref{eq:I3}, respectively)
and $d\Omega$ to the solid angle of the PMT (equation \ref{eq:solid-angle}).
The term $1/\lambda_{s}$ corresponds to the probability for the scattering of the light per unit length. 
The derivative of the time is given by equation \ref{eq:dt/du}.
The distance traveled by the photons is defined as $d = u + v$ (see above).
\newline

The lower and upper limit for the integral of $z$ can be determined 
using equation \ref{eq:z12}.
Apart from the orientation of the PMT, equation \ref{eq:f6} exhibits 
a mirror symmetry in the $x-z$ plane.
One can thus integrate $\phi_0$ between $0$ and $\pi$ and 
evaluate for each $\phi_0$ the angle of incidence of the PMT twice 
using equation \ref{eq:PMT}, 
i.e.\ substituting $\phi_0$ and $-\phi_0$ in equation \ref{eq:vhat}.

\section{Numerical computation}
\label{Numerical computation}

The PDF of the direct light from a muon can be computed directly using equation \ref{eq:f0}.
The PDFs of light from showers and scattered light from a muon 
involves a (multi-dimensional) integral.
In order to evaluate these integrals accurately, 
designated variable transformations are introduced.
\newline

The integral over $\lambda$ is evaluated by integrating over the inverse of the relative group velocity of light ($n_{g}$)
and finding the corresponding wavelength at each point.
As a result, the sampling is denser when the dependence of arrival time on the wavelength 
(i.e.\ the dispersion of light) is stronger.
\newline

The integral over $z$ is evaluated using the variable, $x$:

\begin{eqnarray}
  x & \equiv & e^{\textstyle -a (z - z_2)}   \label{eq:z}
\end{eqnarray}

where the value of the slope parameter, $a$, is set to $a = 1/\lambda_{abs}$.
\newline

The integral over $\phi_0$ in equation \ref{eq:f1} is evaluated using the variable, $y$: 

\begin{eqnarray}
  y & \equiv & e^{\textstyle -b \phi_0}      \label{eq:phi}
\end{eqnarray}

The value of the slope parameter, $b$, is defined as:

\begin{eqnarray}
  b & = & \frac{1}{\pi} \log\frac{v^2_\pi}{v^2_0}
\end{eqnarray}

where 
$v_\pi$ represents the longest possible path length of the photon after the scattering ($\phi_0 \simeq \pi$) and
$v_0$ the shortest possible path length ($\phi_0 \simeq 0$).
The two path lengths can be expressed as:

\begin{eqnarray}
  v_\pi  &  =  & l                                                                        \\
  v_0    &  =  & d - \frac{1}{2} \frac{(d+l) \times (d-l)}{d - l\cos(\theta_C-\theta)}  
\end{eqnarray}

where 
$l = \sqrt{z^2 + R^2}$ and $\theta = \arctan(-R/z)$.
The distance $d$ is defined in equation \ref{eq:d}.
For $z$ close to the position of the \v{C}erenkov cone ($\theta \simeq \theta_C$),
the expression for the shortest possible path length reduces to $v_0 \simeq (d-l)/2$. 
For $\Delta t \simeq 0\;\mathrm{ns}$, 
the $\phi$ dependence of the PDF is expected to be very strong.
The value of the slope parameter is then correspondingly large.
For  $z$ far away from the position of the \v{C}erenkov cone or large $\Delta t$,
the $\phi$ dependence of the PDF is expected to be rather weak.
The value of the slope parameter is then correspondingly small.
\newline

\begin{figure}[ht!]
\setlength{\unitlength}{1cm}
\begin{center}
\begin{picture}(8,8)
\allinethickness{0.35mm}%
\put(1,0){\scalebox{0.8}{%
\put( 1.00, 0.00){\vector(0,1){9}}%
\put( 1.00, 9.50){\makebox(0,0){$z$}}%
\put( 1.00, 8.00){\vector(1,0){8.0}}%
\put( 9.50, 8.00){\makebox(0,0){$x$}}%
\put( 1.00, 1.00){\path(0,0)(3.50,3.50)}%
\put( 4.75, 4.75){\path(0,0)(3.25,3.25)}%
\put( 1.00, 1.00){\circle*{0.2}}%
\put( 0.00, 1.00){\makebox(0,0){$(0,0,z)$}}%
\put( 8.00, 8.00){\circle*{0.2}}%
\put( 8.00, 8.50){\makebox(0,0){$(R,0,0)$}}%
\put( 1.00, 8.00){\circle*{0.2}}%
\put( 0.00, 8.00){\makebox(0,0){$(0,0,0)$}}%
\thinlines%
\put( 1.00, 1.00){\path(0.0,0.0)(2.5,5.0)}%
\put( 3.50, 6.00){\dottedline{0.1}(0.0,0.0)(0.8,1.6)}%
\put( 4.90, 4.90){\makebox(0,0){\rotatebox{135}{\scalebox{5}[1]{\circle{0.7}}}}}%
\put( 5.80, 4.05){\vector(-1,+2){0.1}}%
\put( 3.50, 6.00){\path(0.0,0.0)(4.5,2.0)}%
\put( 2.00, 4.10){\makebox(0,0){$u$}}%
\put( 5.50, 7.40){\makebox(0,0){$v$}}%
\put( 1.00, 1.00){\qbezier(0.6,1.2)(0.82,1.15)(0.92,0.92)}%
\put( 2.20, 2.70){\makebox(0,0){$\alpha$}}%
\put( 3.50, 6.00){\qbezier(0.4,0.8)(0.7,0.7)(0.8,0.4)}%
\put( 4.60, 7.10){\makebox(0,0){$\theta_s$}}%
\put( 8.00, 8.00){\qbezier(-1.2,-0.6)(-1.15,-0.82)(-0.92,-0.92)}%
\put( 6.20, 6.80){\makebox(0,0){$\beta$}}%
\put( 6.30, 4.20){\makebox(0,0){$\phi$}}%
}}%
\end{picture}
\end{center}
\caption{
Definition of the integration angles, $\beta$ and $\phi$, used in the numerical computation of equation \ref{eq:f6}.
\label{f:f3-integral}}%
\end{figure}
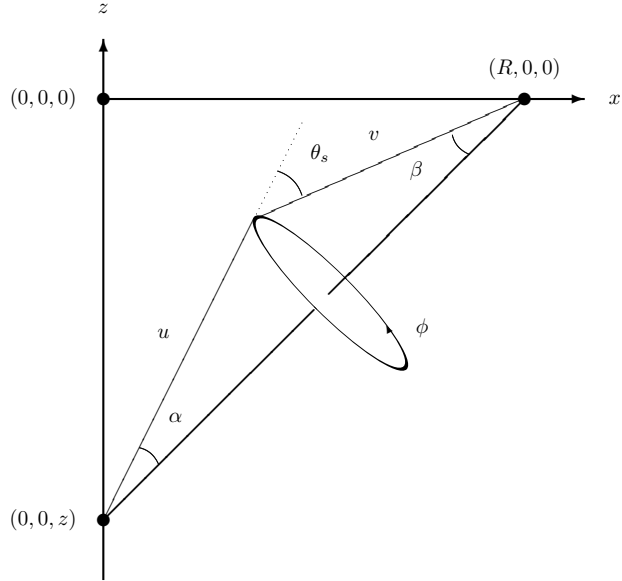

The integral over $\cos\theta_0$ and $\phi_0$ in equation \ref{eq:f6} is 
evaluated using the variables $\sin\beta$ and $\phi$ instead.
These variables are defined in figure \ref{f:f3-integral}.
The integral over $\cos\theta_0$ and $\phi_0$ can then be expressed as:

\begin{eqnarray}
  d\!\cos\theta_0 d\phi_0 &  =  & d\!\cos\alpha d\phi    \\
  \mbox{}                 &  =  & \frac{v^2}{u^2} \tan\beta \; d\!\sin\beta d\phi \label{eq:jacobian}
\end{eqnarray}

The Jacobian in equation \ref{eq:jacobian} 
compensates completely the $v^{-2}$ term in the expression of the PDF due to the solid angle of the PMT (equation \ref{eq:solid-angle}) and
compensates partly     the $(1 - \cos\theta_s)^{-1}$ term due to the time derivative (equation \ref{eq:dt/du}). 
This ensures a proper sampling of the phase-space 
(starting from the muon trajectory it is very unlikely to find the right $\cos\theta_0$ and $\phi_0$
that produce a hit on the PMT at a given time, in particular when the PMT is looking away from the muon).
\newline

Finally, the integrals are evaluated using the Gauss-Legendre technique \cite{ref:numerical-recipes}.
As a result of the variable transformations,
the integrals in equations \ref{eq:f3}, \ref{eq:f1} and \ref{eq:f6} can be evaluated rather accurately
with a relatively small number of integration points (typically 25 for each variable).
It is obvious but worth noting that for each $\Delta t$,
the determination of the value of the PDF of scattered light from a muon and scattered light due to the energy loss of a muon 
thus take only $25^3$ and $25^4$ steps, respectively.

\section{Example plots}
\label{Example plots}

In the following, some example plots are shown.
For this, it is useful to introduce the time difference, $\Delta t$, 
with respect to the expected arrival time assuming the \v{C}erenkov cone hypothesis, i.e:

\begin{eqnarray}
  \Delta t & \equiv & t - t_0 \label{eq:t0}
\end{eqnarray}

The expected arrival time, $t_0$, is defined as:

\begin{eqnarray}
  t_0 & \equiv & \frac{R \; \tan\theta_C}{c}
\end{eqnarray}

where $c$ is the speed of light in vacuum.
The value of the angle $\theta_C$ corresponds here to a typical value.
The definition of the arrival time $t_0$ corresponds to the shortest optical path
from any point on the muon trajectory to the position of the PMT. \\

Unless stated otherwise, 
the PMT is located at a distance of $50\;\mathrm{m}$,
the angular acceptance and the QE of the PMT are taken from figure \ref{f:pmt} and
the probability density function of the light scattering is set to the one labeled \texttt{p0.0075} in figure \ref{f:Ps}.
The absorption length and the scattering length are set to those shown in figure \ref{f:light}.
The values of the other parameters required for the evaluation of the PDFs are summarised in table \ref{t:parameters}. 

\begin{table}[h!]
  \begin{center}
    \renewcommand{\arraystretch}{1.3}
    \begin{tabular}{|l|c|r@{~}l|}
      \hline
      parameter           & symbol                                & \multicolumn{2}{c|}{value}                      \\
      \hline
      \hline
      photo-cathode area  & $A$                                   & $0.00454$            & $\mathrm{m}^{2}$         \\
      Energy loss         & $a(E)$                                & $0.267$              & $\mathrm{GeV/m}$         \\
      Energy loss         & $b(E)$                                & $3.4 \times 10^{-4}$ & $\mathrm{m}^{-1}$        \\
      shower light        & $\frac{\textstyle dx}{\textstyle dE}$ & $4.7319$             & $\mathrm{m/GeV}$         \\*[1mm]
      \hline
    \end{tabular}
  \end{center}
  \caption{\label{t:parameters} Parameter values}
\end{table}

The PDFs of light due to the energy loss of a muon  have been evaluated for a muon energy fixed at $1\;\mathrm{GeV}$.
Because the number of detectable photons is proportional to the muon energy (see equation \ref{eq:I2.1}),
the results should be scaled by a factor 1000 for a muon with an energy of $1\;\mathrm{TeV}$.
The various orientations of the PMT that have been considered are listed in table \ref{t:orientation} 
(see figure \ref{f:event} for the definition of the quoted angles):
\newline

\begin{table}[h!]
\begin{center}
  \renewcommand{\arraystretch}{1.3}
  \begin{tabular}{|c|c|c@{$\;$}l|}
    \hline
    $\theta_{\wp}$    &   $\phi_{\wp}$   &   \multicolumn{2}{c|}{label} \\
    \hline
    \hline
    $0$               &   $0$            &  $\mathcal{N}$  &   North    \\       
    $\pi/2$           &   $0$            &  $\mathcal{E}$  &   East     \\       
    $\pi$             &   $0$            &  $\mathcal{S}$  &   South    \\       
    $\pi/2$           &   $\pi$          &  $\mathcal{W}$  &   West     \\*[1mm]       
    \hline
  \end{tabular}
\end{center} 
\caption{\label{t:orientation} PMT orientations.}
\end{table}

\begin{figure}[ht!]
\begin{center}
\begin{picture}(20,15)
\allinethickness{0.35mm}%
\put( 4.6,-0.2){\makebox(0,0)[b]{\scalebox{0.40}{\includegraphics{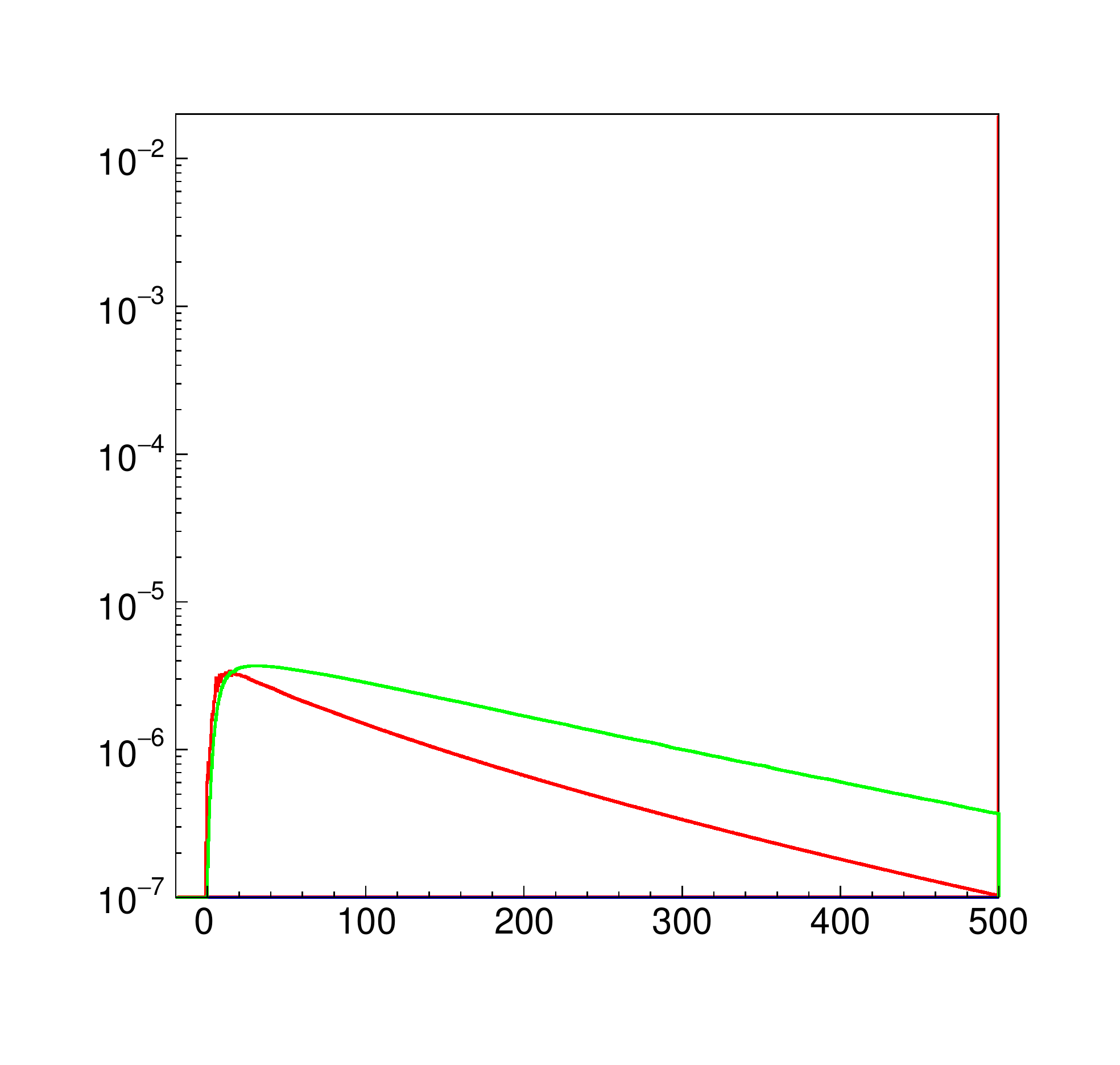}}}}%
\put( 4.8,-0.2){\makebox(0,0){$\Delta t \,[\mathrm{ns}]$}}%
\put( 0.0, 3.6){\makebox(0,0)[l]{\rotatebox{90}{$\frac{\textstyle d\mathcal{P}}{\textstyle dt} ~[\mathrm{p.e./ns}]$}}}%
\put( 2.0, 6.2){\makebox(0,0)[l]{East}}%
\put(12.3,-0.2){\makebox(0,0)[b]{\scalebox{0.40}{\includegraphics{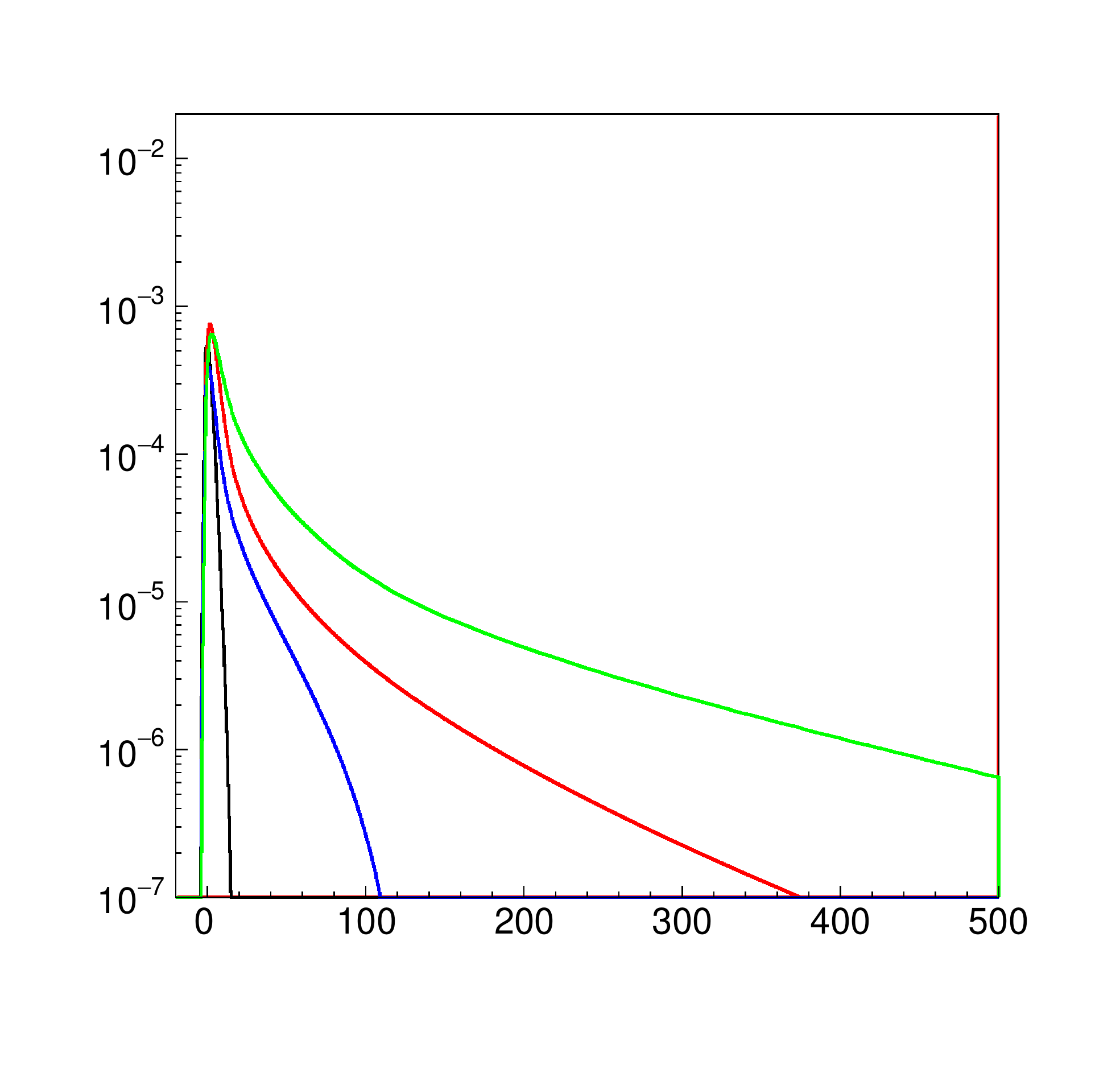}}}}%
\put(12.5,-0.2){\makebox(0,0){$\Delta t \,[\mathrm{ns}]$}}%
\put( 9.8, 6.2){\makebox(0,0)[l]{South}}%
\put(12.3, 7.3){\makebox(0,0)[b]{\scalebox{0.40}{\includegraphics{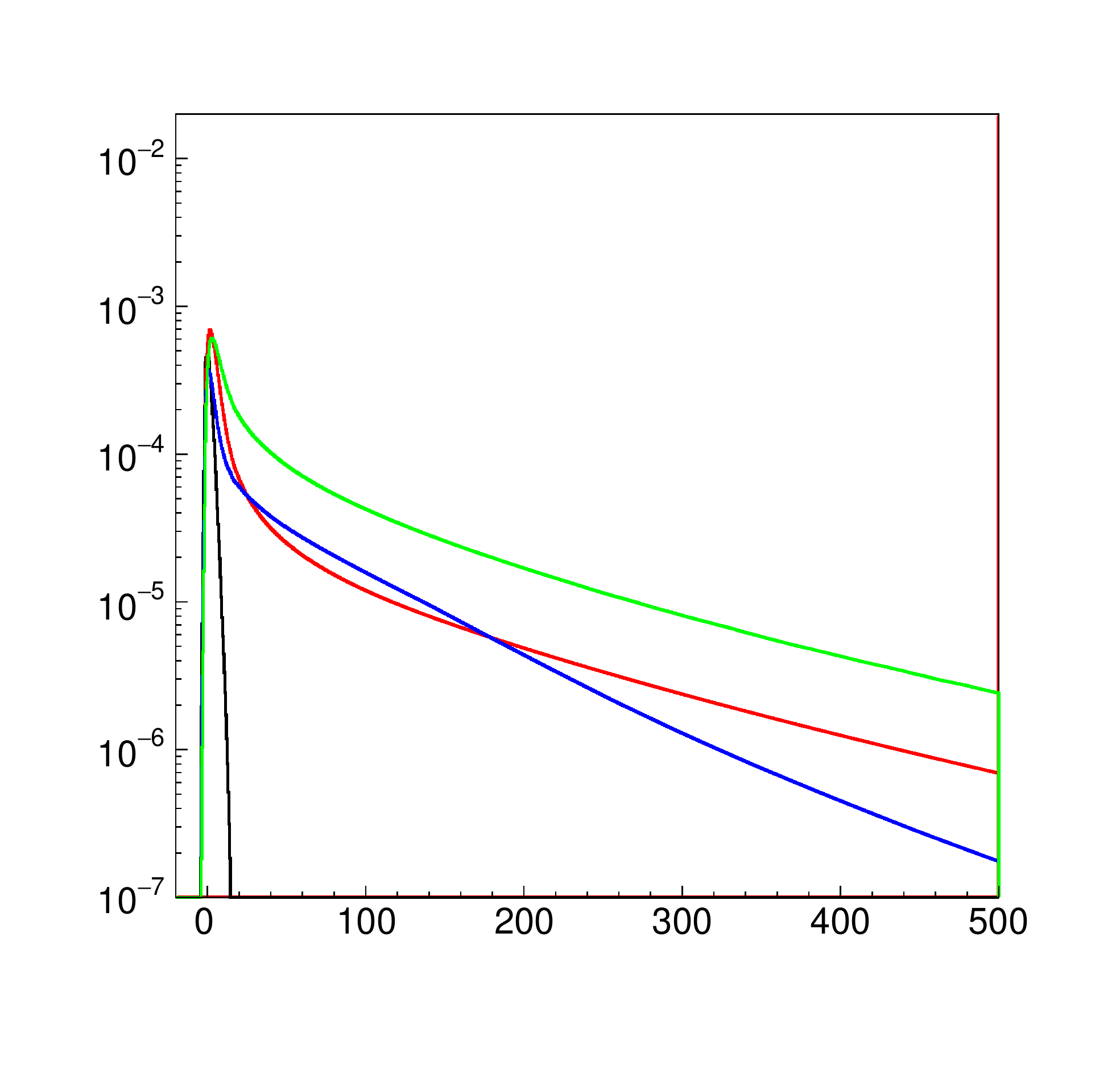}}}}%
\put(12.5, 7.3){\makebox(0,0){$\Delta t \,[\mathrm{ns}]$}}%
\put( 0.0,11.1){\makebox(0,0)[l]{\rotatebox{90}{$\frac{\textstyle d\mathcal{P}}{\textstyle dt} ~[\mathrm{p.e./ns}]$}}}%
\put( 9.8,13.7){\makebox(0,0)[l]{West}}%
\put( 4.6, 7.3){\makebox(0,0)[b]{\scalebox{0.40}{\includegraphics{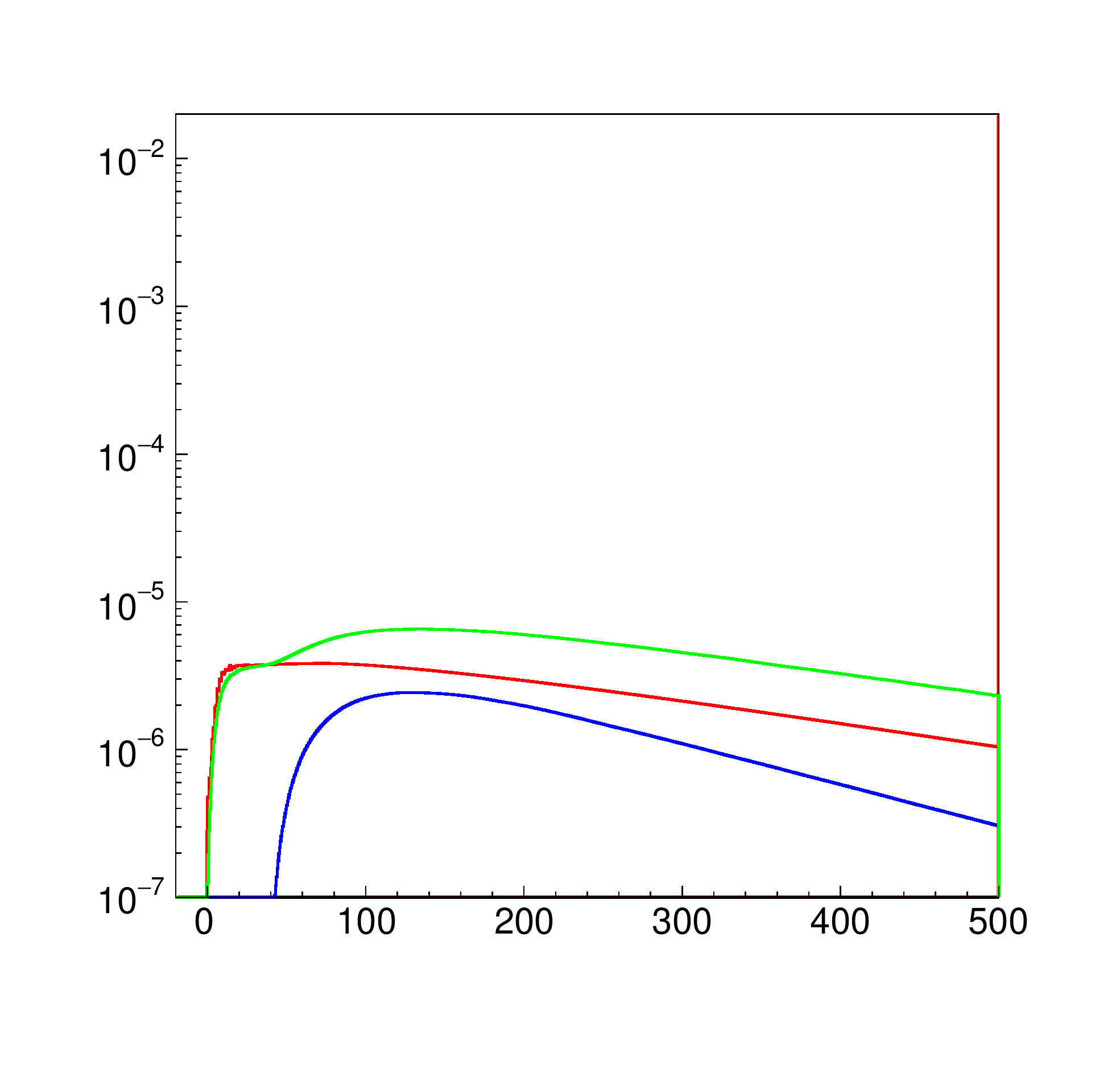}}}}%
\put( 4.8, 7.3){\makebox(0,0){$\Delta t \,[\mathrm{ns}]$}}%
\put( 2.0,13.7){\makebox(0,0)[l]{North}}%
\end{picture}
\end{center}
\caption{
\label{f:pdf} 
PDFs as a function of time for a PMT located at a distance of $50\;\mathrm{m}$.
The labels refer to the PMT orientations listed in table \ref{t:orientation}.
The colour coding is as follows, 
black: muon direct; 
red:   muon indirect;
green: energy loss direct; and 
blue:  energy loss indirect light.
The PDFs for energy loss processes have been normalised to $1\;\mathrm{TeV}$ muon energy. 
}
\end{figure}
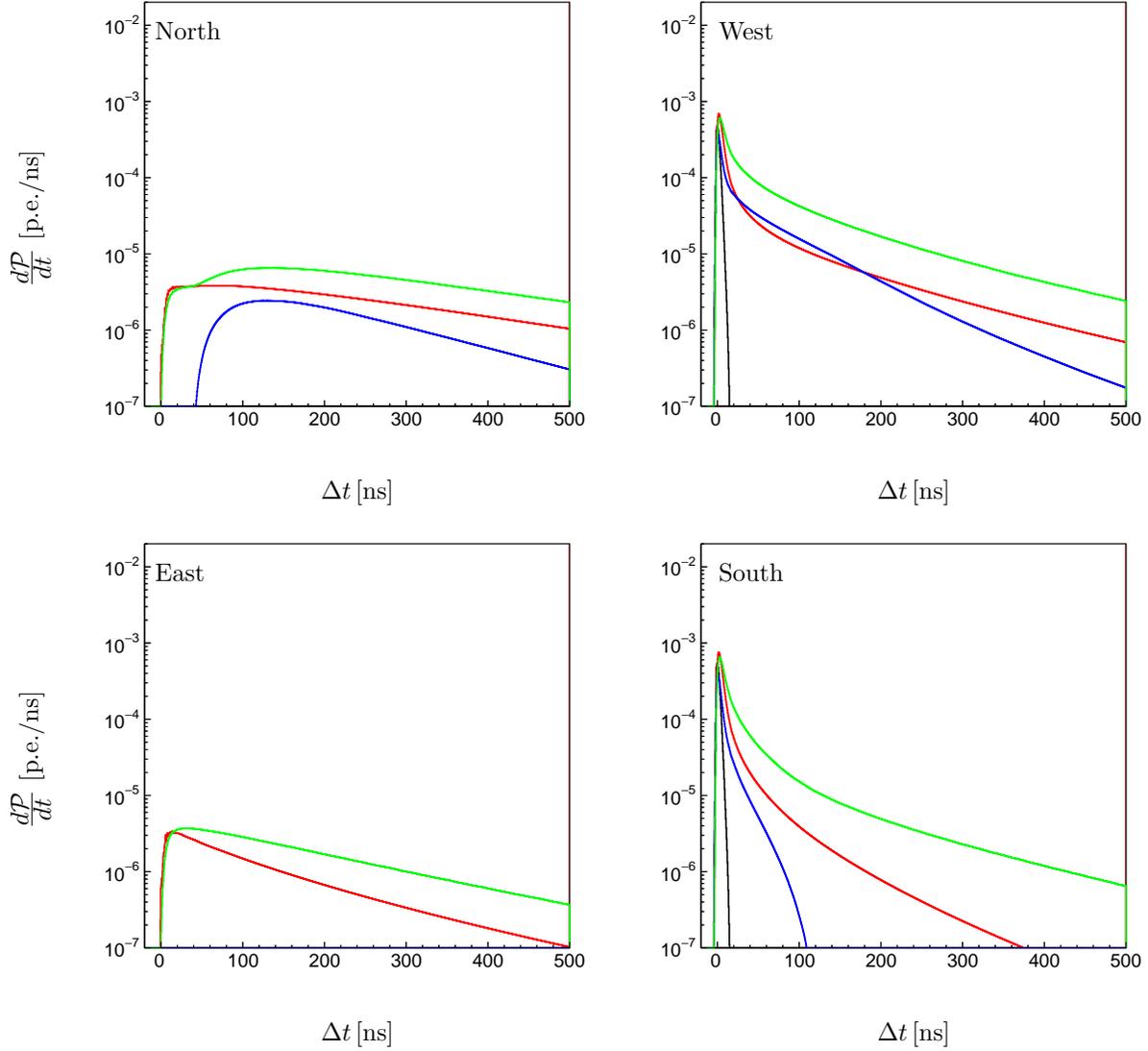

The PDFs as a function of time are shown in figure \ref{f:pdf} for different orientations of the PMT.
As can be seen from figure \ref{f:pdf},
all PDFs exhibit a leading edge at $\Delta t \simeq 0\;\mathrm{ns}$.
For a PMT pointing West or South,
the PDFs show an outstanding peak at the leading edge.
For direct light from a muon, this is conform 
the characteristics of the light cone and 
the definition of $\Delta t$ (equation \ref{eq:t0}).
For direct light from the energy loss processes, this is due to 
the angular distribution of the emitted light which peaks at $\theta_0 = \theta_C$ (see figure \ref{f:geant}).
Under the influence of the scattering of light, the main signature is preserved. 
The preservation of the peak is due to
the shape of the angular distribution of the scattering probability (figure \ref{f:Ps}) and
the pole of the PDF at small $\Delta t$ (equations \ref{eq:dt/dz} and \ref{eq:dt/du}). 
The PDFs of the direct light from the muon for the West and South orientations are identical 
because the angle of incidence of the light on the PMT is the same.
The height of the peaks of the other PDFs are also very similar.
For small $\Delta t$, the optical path is close to that of the \v{C}erenkov cone.
Hence, the PDFs should be very similar indeed because the angles of incidence of the light on the PMT are the same.
The tails of the distributions are, however, quite different.
In general, the light originates from anywhere along the track segment between $z_1$ and $z_2$ (c.f.\ equation \ref{eq:z12}).
Although the track segments are identical in both cases,
due to the absorption of light the downstream end contributes more (c.f.\ equations \ref{eq:absorption} and \ref{eq:d}). 
Consequently, the detected light yield depends on the coverage of the track segment by the field of view of the PMT
which depends on its orientation. 
For a PMT pointing North or East,
the PDFs do not show a significant peak at the leading edge.
No direct light from a muon is detected.
The shape of the PDFs of scattered light from the muon and that due to the energy loss are quite similar.
In general, the PDF for a PMT pointing North is higher than that of a PMT pointing East 
due to the different coverages of the track segment between $z_1$ and $z_2$ by 
the corresponding field of view of the PMT.
This effect is particularly strong for direct light from energy loss processes.
\newline

\appendix

\section{Derivation of $\frac{\partial v}{\partial u}$}
\label{A:derivative}

For the determination of the derivative of the arrival time as a function of the length $u$, 
the quantity $\frac{\partial v}{\partial u}$ should be evaluated.
To this end, equation \ref{eq:position-indirect} should be multiplied with $\hat{u}$.

\begin{eqnarray*}
  Ru_{x} - z u_{z} & = & u + v \cos(\theta_{s})  \\
\end{eqnarray*}

This yields an expression for $v \cos(\theta_{s})$.
Taking the square of equation \ref{eq:position-indirect} yields:

\begin{eqnarray*}
  R^{2} + z^{2}                  & = & u^{2} + v^{2} + 2uv \cos(\theta_{s})                              \\
  \Rightarrow
  R^{2} + z^{2} - u^{2}          & = & v^{2} + 2u \left(Ru_{x} - z u_{z} - u\right)                      \\
  \Rightarrow
  v                             & = & \sqrt{R^{2} + z^{2} + u^{2} - 2u \left(Ru_{x} - z u_{z}\right)}   \\
  \Rightarrow
  \frac{\partial v}{\partial u} & = & -\cos(\theta_{s})
\end{eqnarray*}

\clearpage

\bibliographystyle{plain}
\bibliography{PDF}

\begin{thebibliography}{1}

\bibitem{ref:bailey}
D.\ Bailey.
\newblock ANTARES internal note \verb'ANTARES-PHYSICS-2000-10'.

\bibitem{ref:rayleigh}
C.F. Bohren and D.~Huffman.
\newblock {\em Absorption and scattering of light by small particles}.
\newblock John Wiley, New York, 1983.

\bibitem{ref:kooijman}
P.\ Kooijman.
\newblock private communications.

\bibitem{ref:copper}
C.\ Kopper.
\newblock Performance studies for the km3net neutrino telescope.
\newblock PhD. thesis, University of Erlangen.

\bibitem{ref:mirani}
R.\ Mirani.
\newblock Parametrisation of em-showers in the antares detector volume.
\newblock Doctoral thesis in computational physics, University of Amsterdam.

\bibitem{ref:smith-baker}
Ray Smith and Karen Baker.
\newblock Optical properties of the clearest natural waters (200–800 nm).
\newblock {\em Applied optics}, 20:177--84, 01 1981.

\bibitem{ref:numerical-recipes}
{W.H.~Press, S.A.~Teukolsky, W.T.~Vetterling and B.P.~Flannery}.
\newblock {\em \mbox{NUMERICAL RECIPES IN C++}}.
\newblock Cambridge University Press, 2002.
\newblock {ISBN~0-521-75033-4}.

\bibitem{ref:pdg}
R.~L. Workman et~al.
\newblock {Review of Particle Physics}.
\newblock {\em PTEP}, 2022:083C01, 2022.

\end{thebibliography}

\end{document}